\documentstyle[epsf]{elsart}

\begin{document}
\begin{frontmatter}

\title{Nonlinear Theory of Stochastic Resonance}

\author{Alexander I. Olemskoi\thanksref{AO}}
\address{Sumy State University,
Rimskii-Korsakov St. 2, 40007 Sumy, Ukraine}
\thanks[AO]{E-mail: olemskoi\char'100ssu.sumy.ua}

\begin{abstract}

Theory of nonlinear resonance, including stochastic one, is developed on the basis 
of the statistical field theory and using variables action-angle. 
Explicit expressions of action, proper frequency and nonlinearity parameter 
as functions of the system energy and the external signal frequency are found 
for the cases of nonlinear pendulum and double well potential.

\vspace{0.5cm}


\vspace{0.5cm}


\end{abstract}

\end{frontmatter}

\section{Introduction}

Since the late 1980s, a variety of theoretical and experimental 
papers appears devoting to the stochastic resonance study
and discovering new applications in different fields of science and engineering \cite{5}.
Nowadays, the stochastic resonance is a well established phenomenon displaying in
a bistabile system simultaneously driven by noise and a periodic signal. 
There is appeared to be an optimal noise level at which the system will exhibit 
almost periodic transitions from one state to other with the frequency of the coherent signal. 
The enhancement of a weak input signal  
has been suggested to characterize by the signal-to-noise ratio (SNR) 
that takes a maximum value at an optimal noise level, 
i. e., a behaviour which is reminiscent of a usual resonance phenomenon. 

The observed enhancement is not due to the matching of two frequencies, 
but rather to a cooperative effect of the coherent signal and the noise. 
This peculiarity is reflected formally as follows. 
The kinematic condition
\begin {eqnarray}
\pi r_K=\Omega
\label{A1}
\end {eqnarray}
relates the frequency of the coherent signal $\Omega$
to the Kramers rate $r_K$, which is the function of the noise level 
type of temperature $T$:
\begin {eqnarray}
r_K={\omega_K\over 2\pi}
\exp\left(-{\Delta F\over T}\right),\quad
\omega_K\equiv{\sqrt{F''(x_0)|F''(x_b)|}\over m\gamma}.
\label{A2}
\end {eqnarray}
Here, prime denotes the derivation with respect to a generalized coordinate $x$,  
$F(x)$ is a related dependence of the effective energy, 
which has a minimum at the point $x_0$, a maximum at $x_b$ 
and a barrier of the height $\Delta F\equiv F(x_b)-F(x_0)$;
$m$ is a particle mass, $\gamma$ is a kinetic coefficient. 
Physically, the kinematic condition means that 
for every frequency $\Omega$, the stochastic system could picks out 
suitable noise level $T$. 
However, the value SNR \cite{5}
\begin {eqnarray}
SNR\propto T^{-2}r_K(T),
\label{A3}
\end {eqnarray}
being found on the basis of the statistical theory, 
displays the stochastic resonance maximum at the temperature
\begin {eqnarray}
T_m={1\over 2}\Delta F,
\label{A5}
\end {eqnarray}
which value is fixed by the energy barrier $\Delta F$,   
but not the external signal frequency $\Omega$.

It is easily to see that the reason of this contradiction is using 
the linear approach for the SNR determination. 
In this paper, we develop the theory of nonlinear resonance, including the stochastic one. 
The main ingredients of our approach are:  
(i) the noise accounting by means of introduction of a generalized momentum and 
(ii) passage to the variables action-angle, 
being usual at studying nonlinear phenomena \cite{3}.
Section 2 is devoted to the first of these approaches 
on the basis of the statistical field theory \cite{2}. 
Section 3 contains details of examination of nonlinear resonance in terms of variables action-angle. 
The latters are tested on the example of the simplest model of harmonic oscillator. 
Then, explicit expressions of action, proper frequency and nonlinearity parameter 
as functions of the system energy are found for the cases of nonlinear pendulum, double well potential and
nonlinear pendulum under constant external field. 
Assuming the proper frequency be identical to the frequency of external signal, we examine 
resonance conditions of the nonlinear pendulum in Section 4 and the case of the 
stochastic resonance in Section 5. 
Conclusion in Section 6 shows that the above mentioned contradiction is resolved because  
the nonlinear resonance condition fixes the proper frequency as function of the external one, 
but not the noise level, as in the linear case. 

\section{Basic equations}

Let us study a hydrodynamic mode amplitude $x({\bf r}, t)$
which space-time dependence is determined by the Langevin equation
\begin {eqnarray}
\dot x ({\bf r}, t) = \gamma f(x, t)
+\zeta ({\bf r}, t),\label{1}
\end {eqnarray}
which can be interpreted in the Ito sense.
Here the dot indicates the differentiation with respect to
time $t$, ${\bf r}$ is coordinate, $\gamma$ is a kinetic coefficient,
$f(x, t)$ is a force conjugated to the stochastic variable $x$,
and $\zeta ({\bf r},t)$ is a stochastic term having the form of white noise:
\begin {eqnarray}
\langle\zeta ({\bf r},t) \rangle=0,\quad
\langle \zeta({\bf r},t) \zeta(0,0) \rangle =
T \delta ({\bf r}) \delta(t),
\label{2}
\end {eqnarray}
where the angle brackets
denote averaging and $T$ is a noise intensity type of temperature.
\footnote {The accepted choice of
the coefficient in correlator (\ref{2}) makes it possible to preserve the standard
form of the Focker-Planck equation in which the diffusion term appears with
the coefficient $1/2$ (accordingly, the Gaussian distribution has the standard
dispersion). The doubling of the coefficient leads to the standard Onsager form
of the diffusion component of the flux probability.}
For the system under consideration, the total force
\begin {eqnarray}
f(x, t)=f_0(x)+f_{ext}(t)
\label{3}
\end {eqnarray}
consists of usual internal term
\begin {eqnarray}
f_0(x({\bf r}, t))=-{\delta {\cal F}\over \delta x ({\bf r}, t) },\quad
{\cal F}\{x({\bf r}, t)\}=\int \left[ F(x) +
{\beta\over 2} |\nabla x|^2 \right] {\rm d} {\bf r},
\label{4}
\end {eqnarray}
where $F(x)$ is a system potential per unit
volume, $\beta > 0$ is a constant, $\nabla\equiv\partial/\partial {\bf r}$,
and external harmonic term
\begin {eqnarray}
f_{ext}(t)=A\cos(\Omega t + \varphi)
\label{5}
\end {eqnarray}
being determined by an amplitude $A$, a frequency $\Omega$,
and an initial phase $\varphi$.
It is now convenient
to go over to dimensionless quantities by referring the coordinate ${\bf r}$ to
a characteristic spacing $a$,
the time $t$ and the inverse frequency $\Omega^{-1}$
to the scale $a^3/T\gamma$,
the amplitude $A$, the internal force $f_0$,
and the quantity $F$ to $T/a^3$, and
the fluctuation $\zeta$ to $T\gamma/a^3$.
In this case, Eq.(\ref{1}) can be written as follows:
\begin
{eqnarray} &\dot x = [\nabla^2 x + f(x, t)] + \zeta(t),\quad
f(x, t) = f_0(x) + f_{ext}(t);&\nonumber\\
\label{6}\\
& f_0\equiv -{\partial F}/{\partial x},\quad
f_{ext}(t)=A\cos(\Omega t + \varphi).&\nonumber
\end{eqnarray}
The range of applicability of the Ginzburg-Landau approximation (\ref{4})
is determined by the condition according to which the scale $a$ is much
smaller than the correlation length $\xi=\beta^{1/2}|\partial^2 F/\partial
x^2|^{-1/2}_{x=0}$ \cite{1}. Averaging Eq.(\ref{1})
and disregarding correlations, we
obtain the Landau-Khalatnikov equation for the order parameter
$\left< x ({\bf r},t)\right>$.

The standard field scheme \cite{2} is based on the investigation of the
generating functional corresponding to the stochastic equation (\ref{6}).
It is a functional Laplace transform
\begin{equation}
\label{7}Z\left\{u({\bf r},t) \right\}=\int Z\left\{x({\bf r},t)
\right\}\exp\left(\int u x {\rm d}{\bf r}{\rm d}t \right)Dx ({\bf r},t)
\end{equation}
for the generalized partition function
\begin{equation}
\label{8}Z\left\{x({\bf r},t) \right\}=\left \langle \prod_{({\bf r}
,t)}\delta \left\{ \dot x({\bf r},t)- \nabla^2x ({\bf r},t)-f ({\bf r},t)
-\zeta({\bf r},t) \right\}\det \left| {\frac{{\delta \zeta({\bf r},t)}}{{
\delta x({\bf r},t)}}} \right| \right \rangle_{\zeta}.
\end{equation}
Here, the argument of the $\delta$-function reduces to the Langevin equation
(\ref{6}), and the determinant, providing the passage from the continual
integration over $\zeta({\bf r},t)$ to $x ({\bf r},t)$, is equal to the
unity within the Ito calculus.

In the framework of the standard approach \cite{2}, the $n$-fold variation of
the functional (\ref{7}) with respect to the auxiliary field  $u({\bf r},t)$
allows one to find the $n$-th order correlator for the hydrodynamic
mode amplitude $x({\bf r},t)$ and to construct the perturbation theory.
However, we shall proceed from expression (\ref{8}) for the conjugated
functional $Z\{x({\bf r},t)\}$, variation of which leads to the most
probable realization of the stochastic field $x({\bf r},t)$. Obviously, in
the framework of the mean-field approximation the functional (\ref{8})
reduces to the dependence $Z\{\left\langle x({\bf r},t)\right\rangle \}$,
which corresponds to the Landau free energy
$F\{\left\langle x({\bf r},t)\right\rangle \}=
-T\ln Z\{\left\langle x({\bf r},t)\right\rangle \}$ \cite{1}.

Passing to the consideration of the functional (\ref{8}), we represent the
$\delta $-function in the integral form
\begin{equation}
\label{9}\delta \left\{ x({\bf r},t)\right\} =\int\limits_{-i\infty
}^{i\infty }\exp \left( -\int px{\rm d}{\bf r}{\rm d}t\right) Dp.
\end{equation}
Then, averaging over the noise $\zeta $ with using the Gauss distribution
\begin{equation}
\label{10}P_0\{\zeta \}\propto \exp \left\{ -{\frac 1{{2}}}\int \zeta ^2(
{\bf r},t){\rm d}{\bf r}{\rm d}t\right\} ,
\end{equation}
which corresponds to the condition (\ref{2}), 
and taking into account Eq.(\ref{9}), we
reduce the functional (\ref{8}) to the standard form
\begin{equation}
\label{11}Z\left\{ x({\bf r},t)\right\} =\int P\{x({\bf r},t),p({\bf r}
,t)\}Dp,\quad P\equiv e^{-S}.
\end{equation}
Here, probability distribution $P\{x,p\}$ is given by action $S=\int
{\cal L}{\rm d}{\bf r}{\rm d}t$, where Lagrangian is
\begin{equation}
\label{12}{\cal L}=p\left( \dot x-\nabla ^2x-f\right) -p^2/2.
\end{equation}

Further, we use Euler equations
\begin{equation}
\label{13}{\frac{{\partial {\cal L}}}{{\partial{\rm x}}}}
-{\frac{{\rm d}}{{{\rm d}t}}}
{\frac{{\partial {\cal L}}}{{\partial \dot {\rm x}}}}-
\nabla {\frac{{\partial {\cal L}}}{{\partial \nabla {\rm x}}}}
+\nabla^2 {\frac{{\partial {\cal L}}}{{\partial \nabla^2 {\rm x}}}}=
{\frac{{\partial {\cal R}}}{{\partial \dot {\rm x}}}},\quad
{\rm x}\equiv\{x,p\},
\end{equation}
where dissipative function is
${\cal R}={1\over 2}\dot x^2$.
As a result, equations for the most probable realizations
of the stochastic fields $x({\bf r}, t)$, $p({\bf r},t)$ take the form:
\begin {eqnarray}
&&\dot x=(\nabla^2 x+f)+p,
\label{14}\\
&&\dot p= -\nabla^2 p - f'p - \dot x,
\label{15}
\end{eqnarray}
where prime stands for derivation with respect to variable $x$.
Comparison of (\ref{14}) with the stochastic equation (\ref{6}), having the
same form, shows that the field $p({\bf r}, t)$
is the most probable value of the
fluctuations $\zeta({\bf r}, t)$ of the conjugate force.
Differentiating Eq.(\ref{14}) with respect to the time and inserting result 
into Eq.(\ref{15}), we obtain the equation of motion as follows:
\begin {eqnarray}
\ddot x + (1+f')\dot x = 2f'\nabla^2 x+ \dot f + ff',
\label{16}
\end{eqnarray}
where only terms of the lowest order of spatiotemporal derivations
are kept.

So, usage of above field theory allows us to pass from the stochastic
equation of motion (\ref{1}) of the first order of differentiation 
to the equivalent system
of two equations (\ref{14}), (\ref{15}) of the same order,
or to the single equation (\ref{16}) having the second
differentiation order.  Further, we need in using Hamiltonian ${\cal
H}=p\dot x - {\cal L}$ that depends on the field variable $x$ and the
conjugate momentum $p$.  According to Eq.(\ref{12}), Hamiltonian
can be written in the form 
\begin{eqnarray} 
&{\cal H}(x, p; t)={\cal H}_0(x, p)
+ {\cal H}_1(p, t);&\nonumber\\ \label{17}\\ &{\cal H}_0 = -\nabla
x\nabla p +{1\over 2}p^2+pf_0,\quad {\cal H}_1=Ap\cos(\Omega t +
\varphi).&\nonumber \end{eqnarray} It is easy to see that these
expressions, being inserted into dissipative Hamilton equations
\begin{eqnarray}
\dot x = {\frac{\partial {\cal H}}{\partial p}}
- \nabla{\frac{\partial {\cal H}}{\partial \nabla p}},\quad
\dot p = -\left({\frac{\partial {\cal H}}{\partial x}}
- \nabla{\frac{\partial {\cal H}}{\partial \nabla x}}\right) -
{\frac{{\partial {\cal R}}}{{\partial \dot x}}},
\label{18}
\end{eqnarray}
arrive at the equations of motion (\ref{14}), (\ref{15}).

It is very important to take into account further
that the most probable amplitude $p$ of fluctuations
varies near the magnitude $-f_0$, so that
we ought to pass to oscillating momentum: $p+f_0\to p$.
Moreover, as is seen from the Fourier transformation (\ref{9}),
the momentum $p$ takes imagine magnitudes, so that the power
$f_0$ has to consider as imagine one, as well,
and the sign in front of the last
term of ${\cal H}_0$ in Eqs.(\ref{17}) must be reversed.
\footnote{Apparently, such a reverse corresponds to the passage
from above used Euclidean field theory to the usual one.}
As a result, Hamiltonian (\ref{17}) takes the final form
\begin{eqnarray}
&{\cal H}(x, p; t)={\cal H}_0(x, p) + {\cal H}_1(p, t);&\nonumber\\
\label{17a}\\
&{\cal H}_0 = {1\over 2}p^2 + {1\over 2}f_0^2,\quad
{\cal H}_1=-Af_0\cos(\Omega t + \varphi),&\nonumber
\end{eqnarray}
where gradient terms are suppressed for brevity.

\section{Nonlinear resonance in terms of variables action-angle}

To analyze the system of equations (\ref{14}), (\ref{15}),
it is convenient to use the phase portrait method.
However, in our case such a portrait flows
in the course of the time due to appearance of the time-dependent
external force (\ref{5}).
To avoid this time-variation in phase portrait we need in passage
from above used variables $x$, $p$ to new ones -- action $I$ and
angle $\vartheta$ defined as follows \cite{3}:
\begin{eqnarray}
&I({\cal H})\equiv{1\over 2\pi}\oint p(x, {\cal H}){\rm d}x,&\nonumber\\
\label{19}\\
&\vartheta\equiv{\frac{\partial S(x, I)}{\partial I}},\quad
S(x, I)\equiv\int\limits_0^x p(x', {\cal H}(I)){\rm d}x',&\nonumber
\end{eqnarray}
where the shorted action $S(x, I)$ plays a role
of the generalize function.
The convenience of the so-introduced variables is that the zero Hamiltonian
${\cal H}_0$ in Eqs.(\ref{17a}) does not depend on the angle $\vartheta$,
so that the corresponding phase portrait is stable in the course of the time.
The equations of motion for the generalized coordinate $\vartheta$ and
the conjugate momentum $I$ read (cf. Eqs.(\ref{18}))
\begin{equation}
\dot \vartheta = {\frac{\partial {\cal H}}{\partial I}},\quad
\dot I = -{\frac{\partial {\cal H}_1}{\partial \vartheta}},
\label{20}
\end{equation}
where non-homogeneity and dissipation effects are suppressed.
According to the second of these equations the action $I$
is a constant if an external perturbation is absent.

\subsection{Harmonic oscillator}

To recall a physical meaning of the variables $\vartheta$, $I$ introduced,
let us consider firstly the simplest case of harmonic oscillator.
In this case, the internal power in Eq.(\ref{17a})
\begin{eqnarray}
f_0=-\omega_0 x
\label{21}
\end{eqnarray}
is linear to be fixed by a proper frequency $\omega_0$.
Then, the Hamilton equations (\ref{18}) arrive at the equation
of damping oscillation under external power: 
\begin {eqnarray}
\ddot x + \dot x + \omega_0^2 x =
- A\omega_0\cos(\Omega t + \varphi).
\label{22}
\end{eqnarray}
This equation differs crutially from Eq.(\ref{16}) because of the former 
corresponds to the momentum origin $p=0$, whereas the latter -- to $p=-f_0$. 
According to Eq.(\ref{22}), dissipation shifts resonance frequency
from proper magnitude $\omega_0$ to value 
\begin{eqnarray}
\varpi=\sqrt{\omega_0^2-2^{-2}},
\label{22a}
\end{eqnarray}
whereas a maximum real part of characteristic relation $x/A$
relates to frequency 
\begin{eqnarray}
\omega_{max}=\sqrt{\omega_0^2 - 2^{-1}}.
\label{22b}
\end{eqnarray}
Such a character of the dissipation influence keeps at accounting for
anharmonicity effects if under parameter $\omega_0$ one means 
a proper frequency $\omega({\cal H})$ of nonlinear oscillations 
depending on the system energy.

To demonstrate advantages of using variables action-angle, let us
calculate now their magnitudes at condition that
external power in Hamiltonian (\ref{17a}) is switched off.
Then, the first Eqs.(\ref{19}), (\ref{20}) give immediately
\begin{eqnarray}
I={{\cal H}\over\omega_0},\quad
\dot\vartheta=\omega_0.
\label{23}
\end{eqnarray}
Respectively, the shorted action and angle take the form:
\begin{eqnarray}
&S=I\left[{\rm arcsin}\left({x\over x_0}\right)
+{x\over x_0}\sqrt{1-\left({x\over x_0}\right)^2}\right];&\nonumber\\
\label{24}\\
&\vartheta={\rm arcsin}\left({x\over x_0}\right),\quad
x_0^2\equiv{2I\over\omega_0}.&\nonumber
\end{eqnarray}
The last of Eqs.(\ref{23}) gives the usual relation
between the angle and the time
\begin{eqnarray}
\vartheta=\omega_0 t + \vartheta_0,
\label{25}
\end{eqnarray}
which using arrives at the harmonic laws of motion  
\begin{eqnarray}
&&x=(2{\cal H}/\omega_0^2)^{1/2}{\rm sin}(\omega_0 t + \vartheta_0),\nonumber\\
&&p=(2{\cal H})^{1/2}{\rm cosin}(\omega_0 t + \vartheta_0).
\label{25aa}
\end{eqnarray}

\subsection{Nonlinear pendulum}

The simplest model of system with possibility of the barrier overcoming 
is known to be the nonlinear pendulum 
(in this Subsection, we consider the pendulum without a friction and
an external perturbation).
Here, Hamiltonian takes the form
\begin{eqnarray}
{\cal H}_0 = {1\over 2}p^2 + 2\omega_0^2\sin^2(x/2),
\label{26}
\end{eqnarray}
corresponding to the power $f_0=-2\omega_0\sin(x/2)$ in Eqs.(\ref{17a}).
Hamilton equations (\ref{18}) arrive at the system
\begin{eqnarray}
\dot x = p,\quad
\dot p = - \omega_0^2\sin x.
\label{27}
\end{eqnarray}
Combination of these equations gives the nonlinear one:
\begin{eqnarray}
\ddot x + \omega_0^2\sin x = 0.
\label{28}
\end{eqnarray}
This equation is non-solvable in analytical form and
we ought to use the phase portrait method.
The form of this portrait, following from Eqs.(\ref{27}),
shows that the system behaviour
is governed by energy ${\cal H}$ with respect to the critical value
${\cal H}_c\equiv 2\omega_0^2$.
At condition ${\cal H}<{\cal H}_c$, the system moves finitely, whereas
with overcoming the critical energy ${\cal H}_c$ it passes to
infinite motion.
Let us describe such a behaviour quantitatively.

In this line, the simplest topic is the solution
corresponding to separatrix, for which the energy is critical one:
${\cal H}={\cal H}_c$. In such a case, the definition (\ref{26}) gives
the separatrix form as follows:
\begin{eqnarray}
p=\pm 2\omega_0\cos(x/2).
\label{29}
\end{eqnarray}
Then, the first of Eqs.(\ref{27}) arrives at the separatrix law of motion 
\begin{eqnarray}
x=4{\rm arctan}\exp(\pm\omega_0 t)-\pi,
\label{30}
\end{eqnarray}
where the different signs correspond to upper and lower branches
of the separatrix. This dependence can be written
in much more elegant form $\cos(x/2)=[{\rm cosh}(\pm\omega_0 t)]^{-1}$,
insertion of which into Eq.(\ref{29}) arrives at the famous
soliton dependence       
\begin{eqnarray}
p=\pm {2\omega_0\over{\rm cosh}(\omega_0 t)},
\label{31}
\end{eqnarray}
where choice of signs corresponds to solitons moving to right or left
sides.  In accordance with the motion laws (\ref{30}), (\ref{31}),
the system behaviour on the separatrix (\ref{29}) is as follows:
at time $t=-\infty$ the system is located in a saddles $S_{\mp}$, where
the coordinate $x=\mp\pi$ and the momentum $p=0$.
In the course of the time within the domain $-\infty<t<\infty$,
the former arises monotonously from $-\pi$ to $\pi$,
whereas the latter increases at $t<0$ and decreases at $t>0$. 
It is characteristically that coordinate variation and 
finite magnitudes of the momentum take place within the domain
$\Delta t\sim\omega_0^{-1}$ located near the time $t=0$.
The forms of the corresponding kink $x(t)$ and soliton $p(t)$ 
are depicted in Fig.~2.

General solution of Eqs.(\ref{27}) can be obtained with using
the variables action-angle defined by Eqs.(\ref{19}).
It is convenient to introduce a parameter
\begin{eqnarray}
\kappa^2\equiv{1\over 2}{{\cal H}\over\omega_0^2}\equiv
{{\cal H}\over{\cal H}_c},\quad {\cal H}_c\equiv 2\omega_0^2,
\label{32}
\end{eqnarray}
taking the magnitude $\kappa=1$ at critical energy ${\cal H}={\cal H}_c$,
and a new variable $\xi$ defined by equalities
\begin{eqnarray}
\sin\xi\equiv\left\{\matrix{\kappa^{-1}\sin(x/2)&\qquad{\rm at}
\qquad\kappa\leq 1,\cr
\sin(x/2)&\qquad{\rm at}\qquad\kappa\geq 1.\cr}\right.
\label{33}
\end{eqnarray}                                  
Then, the generating function is expressed in terms of
the incomplete Jacobian elliptic integrals $F(\xi, \kappa)$, $E(\xi,
\kappa)$ of the first and second orders \cite{4} as follows:
\begin{eqnarray}
S(x, I)=4\omega_0\left\{\matrix
{\left[E(\xi, \kappa)-(1-\kappa^2)F(\xi, \kappa)\right]
&\quad{\rm at}\quad\kappa\leq 1,\cr
\kappa E(\xi, 1/\kappa)
&\quad{\rm at}\quad\kappa\geq 1.\cr}\right.
\label{34}
\end{eqnarray}
Differentiation of these equalities with respect to $I$ arrives at expressions
for the angle $\vartheta$ that generalyzes the last equality (\ref{24}) (we
supress these expressions because of its very complicative form).

Fortunately, formulas for action $I\equiv 4S(\xi=\pi/2)/2\pi$
follow from Eqs.(\ref{34}) immediately and are expressed by means of
the complete Jacobian elliptic integrals $K(\kappa)\equiv F(\xi=\pi/2, \kappa)$,
$E(\kappa)\equiv E(\xi=\pi/2, \kappa)$. 
Taking into account corresponding dependences shown in Fig.~3, 
we shall need further in using asymptotics of these integrals \cite{4}
\begin{eqnarray}
K(\kappa)\approx\left\{\matrix{
{\pi\over 2}\left(1+{\kappa^2\over 4}\right)
&\quad{\rm at}\quad\kappa\ll 1,\cr
\ln{4\over\sqrt{1-\kappa^2}}
&\quad{\rm at}\quad 1-\kappa^2\ll 1;\cr}\right.
\label{34a}
\end{eqnarray}
\begin{eqnarray}
E(\kappa)\approx\left\{\matrix{
{\pi\over 2}\left(1-{\kappa^2\over 4}\right)
&\quad{\rm at}\quad\kappa\ll 1,\cr
1+{1-\kappa^2\over 2}\ln{4\over\sqrt{1-\kappa^2}}
&\quad{\rm at}\quad 1-\kappa^2\ll 1.\cr}\right.
\label{34b}
\end{eqnarray}
Resulting dependence $I({\cal H})$ depicted in Fig.~4 shows monotonic increase 
from $I=0$ at ${\cal H}=0$ to infinity with the logarithmical inflection at 
the critical energy ${\cal H}_c$. 
This behaviour is characterized by the following asimptotics: 
\begin{eqnarray}
I\approx 2\omega_0\left\{\matrix{{\cal H}/{\cal H}_c
&\quad {\rm at} \quad{\cal H}\ll{\cal H}_c,\cr
{4\over\pi}\left(1-{1-{\cal H}/{\cal H}_c\over 4}
\ln{16\over 1-{\cal H}/{\cal H}_c}\right)
&\quad {\rm at} \quad 0<{\cal H}_c-{\cal H}\ll{\cal H}_c,\cr
{4\over\pi}\left(1+{{\cal H}/{\cal H}_c-1\over 4}
\ln{16\over {\cal H}/{\cal H}_c-1}\right)
&\quad {\rm at} \quad 0<{\cal H}-{\cal H}_c\ll{\cal H}_c,\cr
2\sqrt{{{\cal H}\over{\cal H}_c}}\left(1-{{\cal H}_c\over 4{\cal H}}\right)
&\quad {\rm at} \quad{\cal H}\gg{\cal H}_c.\cr}\right.
\label{45aa}
\end{eqnarray}

Accounting the properties of the elliptic integrals \cite{4},
we obtain for the proper frequency $\omega\equiv\dot\vartheta$
determined by the first equality (\ref{20}) in the following form:
\begin{eqnarray}
\omega={\pi\over 2}{\omega_0\over K(\bar\kappa)}\left\{\matrix{
1&\qquad {\rm at}\qquad \kappa\leq 1,\cr
\kappa&\qquad {\rm at}\qquad \kappa\geq 1,\cr}\right.
\label{35}
\end{eqnarray}
where
\begin{eqnarray}
&&\bar\kappa\equiv\left\{\matrix{\kappa
&\qquad {\rm at}\qquad \kappa\leq 1,\cr
\kappa^{-1}&\qquad {\rm at}\qquad \kappa\geq 1.\cr}\right.
\label{35a}
\end{eqnarray}
As is seen in Fig.~4, the proper frequency falls down
from the bare magnitude $\omega_0$ at the minimal energy ${\cal H}=0$
to zero at ${\cal H}={\cal H}_c$ and then, after infinitely sharp cusp,
the value $\omega$ increases monotonously.
According to Eqs.(\ref{34a}), such a behaviour is presented
by asimptotics:
\begin{eqnarray}
\omega\approx\omega_0\left\{\matrix{1-{{\cal H}\over 4{\cal H}_c}
&\quad{\rm at}\quad{\cal H}\ll{\cal H}_c,\cr
\pi\left(\ln{16\over|1-{\cal H}/{\cal H}_c|}\right)^{-1}
&\quad{\rm at}\quad |{\cal H}-{\cal H}_c|\ll{\cal H}_c,\cr
\sqrt{{{\cal H}\over{\cal H}_c}}\left(1-{{\cal H}_c\over 4{\cal H}}\right)
&\quad{\rm at}\quad{\cal H}\gg{\cal H}_c.\cr}\right.
\label{36}
\end{eqnarray}
On the other hand, definitions (\ref{26}), (\ref{33})
arrive at time-dependencies of the momentum:
\begin{eqnarray}
p=\pm 2\omega_0\kappa\left\{\matrix{\cos\xi={\rm cn}(t, \kappa)\quad  
{\rm at}\quad {\cal H}\leq{\cal H}_c,\cr
\sqrt{1-\kappa^{-2}\sin^2\xi}={\rm dn}(t, \kappa^{-1})
&\quad {\rm at}\quad {\cal H}\geq{\cal H}_c,\cr}\right.
\label{37}
\end{eqnarray}
where $\kappa^2\equiv{\cal H}/{\cal H}_c$; 
${\rm cn}(t, \kappa)$, ${\rm dn}(t, \kappa^{-1})$ are
the Jacobian elliptic functions shown in Fig.~3.
With accounting for Eqs.(\ref{33}),
these expressions pass to Eqs.(\ref{29}), (\ref{31}) at ${\cal H}={\cal H}_c$.

To elucidate the system behaviour with energy increase,
let us expand the dependences (\ref{37}) into Fourier series \cite{3}
\begin{eqnarray}
p=\pm 8\omega\left\{\matrix{\sum\limits_{n=1}^{\infty}a_n\cos [(2n-1)\omega t]
&\quad{\rm at}\quad{\cal H}\leq{\cal H}_c,\cr
{1\over 4}+\sum\limits_{n=1}^{\infty}a_n\cos (n\omega t)
&\quad{\rm at}\quad{\cal H}\geq{\cal H}_c,\cr}\right.
\label{38}
\end{eqnarray}
where one denotes
\begin{eqnarray}
&a_n\equiv\left\{\matrix{{k^{n-1/2}\over 1+k^{2n-1}}
&\quad{\rm at}\quad {\cal H}\leq{\cal H}_c,\cr
{k^{n}\over 1+k^{2n}}
&\quad{\rm at}\quad{\cal H}\geq{\cal H}_c;\cr}\right.\label{39}\\
&k\equiv\exp\left(-\pi{K'\over K}\right),
\quad K'\equiv K(\sqrt{1-\bar\kappa^2}),
\quad K\equiv K(\bar\kappa),&\nonumber\\
\nonumber
\end{eqnarray}
parameter $\bar\kappa$ is determined by Eq.(\ref{35a}).
According to Eqs.(\ref{34a}), one has asymptotics
\begin{eqnarray}
k\approx\left\{\matrix{\kappa^2/32
&\quad{\rm at}\quad\kappa\ll 1,\cr
k\approx\exp(-\pi/N)
&\quad{\rm at}\quad 1-\kappa^2\ll 1,\cr}\right.
\label{40}
\end{eqnarray}
where number $N\equiv\omega_0/\omega$ is asimptotically as follows:
\begin{eqnarray}
N\approx\left\{\matrix{1
&\quad{\rm at}\quad{\cal H}\ll{\cal H}_c,\cr
{1\over\pi}\ln{16{\cal H}_c\over{|\cal H}-{\cal H}_c|}
&\quad{\rm at}\quad |{\cal H}-{\cal H}_c|\ll{\cal H}_c.\cr}\right.
\label{41}
\end{eqnarray}
Thus, 
at low energies a single harmonics prevails to correspond to 
the coefficient of the Fourier series (\ref{38})
\begin{eqnarray}
a_n\approx\left({{\cal H}\over 32{\cal H}_c}\right)^{n-1/2},
\qquad{\cal H}\ll{\cal H}_c.
\label{42}
\end{eqnarray}
Respectively, near the critical energy, where the Fourier series gets
the harmonics number $N\gg 1$, one obtaines
\begin{eqnarray}
a_n\approx 8\omega\left\{\matrix{1 
&\quad{\rm at}\quad 1<n<N,\cr
\exp[-\pi(n/N)]
&\quad{\rm at}\quad n>N.\cr}\right.
\label{43}
\end{eqnarray}

As is known \cite{3}, above described behaviour is characterized
by nonlinearity parameter
\begin{eqnarray}
\alpha\equiv\left|{{\rm d}\ln\omega\over{\rm d}\ln I}\right|.
\label{44}
\end{eqnarray}
According to (\ref{34}), (\ref{35}), this parameter is determined
by the following equation:
\begin{eqnarray}
\alpha=\left\{\matrix{{1-\kappa^2\over\kappa^2}
\left[{1\over 1-\kappa^2}{E(\kappa)\over K(\kappa)}-1\right]^2
&\quad{\rm at}\quad\kappa\leq 1,\cr
{\kappa^2\over\kappa^2-1}
\left({E(\kappa^{-1})\over K(\kappa^{-1})}\right)^2
&\quad{\rm at}\quad\kappa\geq 1.\cr}\right.
\label{45}
\end{eqnarray}
The nonlinierity parameter takes on value
$\alpha=0$ at ${\cal H}=0$ and then arises indefinitely at the 
critical energy ${\cal H}_c$, tending to magnitude $\alpha=1$ at 
${\cal H}\to\infty$. 
Such a behaviour is characterized by the following asimptotics: 
\begin{eqnarray}
\alpha\approx\left\{\matrix{{1\over 4}({\cal H}/{\cal H}_c)
&\quad {\rm at} \quad{\cal H}\ll{\cal H}_c,\cr
4\left(1-{{\cal H}\over{\cal H}_c}\right)^{-1}
\left(\ln{16\over 1-{\cal H}/{\cal H}_c}\right)^{-2}
\left[1-{1-{\cal H}/{\cal H}_c\over 2}
\ln{16\over 1-{\cal H}/{\cal H}_c}\right]\
&{\rm at} \ 0<{\cal H}_c-{\cal H}\ll{\cal H}_c,\cr
4\left({{\cal H}\over{\cal H}_c}-1\right)^{-1}
\left(\ln{16\over {\cal H}/{\cal H}_c-1}\right)^{-2}
\left[1+{{\cal H}/{\cal H}_c-1\over 2}
\ln{16\over{\cal H}/{\cal H}_c-1}\right]\
&{\rm at} \ 0<{\cal H}-{\cal H}_c\ll{\cal H}_c,\cr
1-{\cal H}_c/{\cal H}
&\quad {\rm at} \quad{\cal H}\gg{\cal H}_c.\cr}\right.
\label{45a}
\end{eqnarray}

As a result, the observed picture of nonlinear oscillation is as follows.
At low energy, when ${\cal H}\ll{\cal H}_c$, only single harmonic with the frequency
$\omega\approx\omega_0$ keeps in the Fourier series (\ref{38}), so that
the low-energy limit reduces to above considered case
of harmonic oscillation (see Subsection 3.1).
With energy increase, the harmonics number $N$ arises in a manner
of the dependence $K(\kappa)$ shown in Fig.~3,
taking logarithmically large magnitudes (\ref{41}) near
the critical value ${\cal H}_c\equiv 2\omega_0^2$.
On the other hand, the oscillation frequancy (\ref{36}) and
the harmonic amplitudes (\ref{43}) decrease monotonously to zero.
Thus, one can mean in coarse manner that, with energy increase
in the domain $0\leq{\cal H}\leq{\cal H}_c$, the single harmonic oscillation
transforms to a harmonics superposition, whose number $N$
increases monotonously to infinity, whereas frequency $\omega$
and amplitude $a_n\sim 8\omega$ decrease to zero.
Just under the critical energy $(0<{\cal H}_c-{\cal H}\ll{\cal H}_c)$
the system behaviour is characterizes by a set of solitons
of different signs, whereas just above ${\cal H}_c$ the signs of these solitons
become equal (see Eqs.(\ref{37})). Remarcable peculiarity
of such solitons set is that the width of a single soliton
is reduced to $\Delta t\sim\omega_0^{-1}$, whereas a distance between them
is $N\Delta t\sim\omega^{-1}\gg\omega_0^{-1}$.
For just critical energy ${\cal H}_c$, the system behaviour is caracterized
by the separatrix solution (\ref{31}) that reduces to the single soliton.

\subsection{Double well potential}

Taking into account use in the stochastic resonance problem \cite{5}, 
let us consider now the oscillation in double well potential.
We will show that, in terms of the variables action-angle,
the difference with above studied case of the nonlinear pendulum 
is quite quantitive, but not qualititive.

The basic model is presented by the potential
\begin{eqnarray}
F={\omega_0^2\over 4}(1-x^2)^2,
\label{46}
\end{eqnarray}
counting off the energy at equilibrium positions $x=\pm 1$,
$\omega_0$ is bare frequency. In this case, Hamilton equations (\ref{18})
take the forms type of Eqs.(\ref{27}):
\begin{eqnarray}
\dot x = p,\quad
\dot p = \omega_0^2 x(1-x^2).
\label{27a}
\end{eqnarray}
Near the saddle point $x=0$, $p=0$, corresponding phase portrait 
has the form differring from the same  
for nonlinear pendulum.
This form is characterized by the separatrix (cf. Eq.(\ref{29}))
\begin{eqnarray}
p=\pm \omega_0 x\sqrt{1-x^2/2},
\label{29a}
\end{eqnarray}
corresponding to the condition ${\cal H}=(\omega_0/2)^2$.
Similarly to the case of nonlinear pendulum, the first of Eqs.(\ref{27a}) 
arrives at the separatrix law of motion:
\begin{eqnarray}
x=\pm 2^{1/2}\sqrt{1-[{\rm sinh}(\omega_0 t)]^2}.
\label{30a}
\end{eqnarray}
For the time-dependence of the momentum, one has double-soliton solution 
(cf. Eq.(\ref{31})) 
\begin{eqnarray}
p=\mp 2^{1/2}\omega_0{\rm sinh}(\omega_0 t)\sqrt{1-[{\rm sinh}(\omega_0 t)]^2},
\label{31a}
\end{eqnarray}
where choice of signs corresponds to solitons moving to right or left sides. 
According to the motion laws (\ref{30a}), (\ref{31a}) -- on the one hand, and 
Eqs.(\ref{30}), (\ref{31}) -- on the other,
the difference between the separatrix solutions for double well potential 
and nonlinear pendulum is that, in the course of the time near the point $t=0$, 
the momentum gains two peaks of different signs in the former case 
and the single peak in the latter. 

To introduce the variables action-angle, it is convenient to use
parameter $\kappa$ type of given by Eq.(\ref{32}) and new variable $\xi$
determined by equality type of (\ref{33}):
\begin{eqnarray}
\kappa^2\equiv
{{\cal H}\over{\cal H}_c},\quad 
{\cal H}_c\equiv\left({\omega_0\over 2}\right)^2;
\label{32b}
\end{eqnarray}
\begin{eqnarray}
x^2\equiv 1-\left\{\matrix{\kappa\sin\xi
&\qquad{\rm at}\qquad\kappa\leq 1,\cr
\sin\xi&\qquad{\rm at}\qquad\kappa\geq 1.\cr}\right.
\label{32aa}
\end{eqnarray}
Moreover, we shall need in using integrals
\begin{eqnarray}
&{\cal I}_n(\alpha, \kappa)\equiv\int\limits_{0}^{\alpha}
{(1-\kappa\xi)^{n-{1\over 2}}\over\sqrt{1-\xi^2}}{\rm d}\xi,
\ \alpha\leq 1, \ n=0, \pm 1, \pm 2,\dots \quad {\rm at}\quad \kappa\leq 1;&
\nonumber\\
\label{33a}\\
&{\cal J}_n(\alpha, \kappa)\equiv\int\limits_{0}^{\alpha}
{(1-\kappa^{-2}\xi^2)^{{1\over 2}-n}\over\sqrt{1-\xi}}{\rm d}\xi,
\ \alpha\leq 1, \ n=0, 1, 2,\dots \quad {\rm at}\quad \kappa\geq 1.&
\nonumber
\end{eqnarray}
The form of dependencies 
${\cal I}_n(\kappa)\equiv{\cal I}_n(\alpha=1, \kappa)$, 
${\cal J}_n(\kappa)\equiv{\cal J}_n(\alpha=1, \kappa)$ is depicted 
in Fig.~6, the corresponding asymptotics read:
\begin{eqnarray}
&{\cal I}_n(\kappa)\approx{\pi\over 2}-{1\over 2}(2n-1)\kappa,\quad \kappa\ll 1;&
\nonumber\\
\label{33d}\\
&{\cal J}_n(\kappa)\approx 2+{8\over 15}(2n-1)\kappa^{-2},\quad \kappa\gg 1.&
\nonumber
\end{eqnarray}
The integrals are subjected to simple derivation rules
\begin{eqnarray}
&\kappa{{\rm d}{\cal I}_n\over{\rm d}\kappa}=
\left(n-{1\over 2}\right)\left({\cal I}_n-{\cal I}_{n-1}\right),&
\nonumber\\
\label{33aa}\\
&\kappa{{\rm d}{\cal J}_n\over{\rm d}\kappa}=
(2n-1)\left({\cal J}_n-{\cal J}_{n-1}\right),&
\nonumber
\end{eqnarray}
where the arguments $\alpha$, $\kappa$ are supressed for brevity.

As a result, the last of definitions (\ref{19})
arrives at the expression (cf. Eq.(\ref{34}))
\begin{eqnarray}
S={\omega_0\over 2\sqrt{2}}\left\{\matrix{
-(1-\kappa^2){\cal I}_0 +2{\cal I}_1 - {\cal I}_2
&\quad{\rm at}\quad\kappa\leq 1,\cr
\kappa{\cal J}_0
&\quad{\rm at}\quad\kappa\geq 1.\cr}\right.     
\label{34c}
\end{eqnarray}
The action $I\equiv(2/\pi)S$ follows from this at $\alpha=1$.
Respectively, for the proper frequency one obtains instead of Eq.(\ref{35})
\begin{eqnarray}
\omega=\sqrt{2}\pi\omega_0\kappa^2\left\{\matrix{
\left[-(1-\kappa^2){\cal I}_{-1} +
(3\kappa^2-1){\cal I}_0 + 5{\cal I}_1 - 3{\cal I}_2\right]^{-1}
&\quad{\rm at}\quad\kappa\leq 1,\cr
\left(2\kappa{\cal J}_1\right)^{-1}
&\quad{\rm at}\quad\kappa\geq 1.\cr}\right.     
\label{35c}
\end{eqnarray}
Energy dependencies $I({\cal H})$, $\omega({\cal H})$, 
following from Eqs.(\ref{34c}), (\ref{35c}), are depicted in Fig.~7.
It is seen that these take the form type of 
the corresponding dependences for nonlinear pendulum (see Fig.~4). 
Fourier spectrum of the time dependence $p(t)$ of the momentum 
behaves in analogous manner as in Eqs.(\ref{38}):
with energy increase in the domain $0\leq {\cal H}\leq {\cal H}_c$,
single harmonic oscillation transforms to a harmonic superposition,
whose number increases monotonously to infinity, whereas frequencies
and amplitudes decrease to zero. 
Such a behaviour is characterized by nonlinearity parameter (\ref{44}), 
taking the following form (cf. Eq. (\ref{45}))
\begin{eqnarray}
\alpha=\left\{\matrix{
6{\left[(1-\kappa^2){\cal I}_{-2} - 2{\cal I}_{-1} +
\kappa^2{\cal I}_0 + 2{\cal I}_1 - {\cal I}_2\right]
\left[-(1-\kappa^2){\cal I}_0 +2{\cal I}_1 - {\cal I}_2\right]
\over\left[-(1-\kappa^2){\cal I}_{-1} +
(3\kappa^2-1){\cal I}_0 + 5{\cal I}_1 - 3{\cal I}_2\right]^{2}}
&\quad{\rm at}\quad\kappa\leq 1,\cr
{{\cal J}_0{\cal J}_2\over{\cal J}_1^2}
&\quad{\rm at}\quad\kappa\geq 1.\cr}\right.     
\label{45a}
\end{eqnarray}
Respectively, double curvature $\Delta\equiv 2{{\rm d}\omega\over{\rm d}I}$ 
of the dependence ${\cal H}(I)$ is connected with the parameter $\alpha$ 
as $\Delta\equiv 2(\omega/I)\alpha$ to read
\begin{eqnarray}
\Delta=2\pi^2\left\{\matrix{
6\kappa^2{(1-\kappa^2){\cal I}_{-2} - 2{\cal I}_{-1} +
\kappa^2{\cal I}_0 + 2{\cal I}_1 - {\cal I}_2
\over\left[-(1-\kappa^2){\cal I}_{-1} +
(3\kappa^2-1){\cal I}_0 + 5{\cal I}_1 - 3{\cal I}_2\right]^{3}}
&\quad{\rm at}\quad\kappa\leq 1,\cr
{{\cal J}_2\over{\cal J}_1^3}
&\quad{\rm at}\quad\kappa\geq 1.\cr}\right.     
\label{45b}
\end{eqnarray}

\subsection{Nonlinear pendulum under constant external field}

Before studying the effect of periodical external field,
let us anounce main peculiatities of the constant perturbation
following supersymmetry theory \cite{6}.
In this case, the potential energy
\begin {eqnarray}
F=2\omega_0^2\sin(x/2) - {\cal E}x
\label{a}
\end {eqnarray}
is characterized, becides the proper frequency $\omega_0$,
by a field strenth ${\cal E}$.  
Swithching such bias field arrives at the expression for flux 
$j\equiv\langle x\rangle/t$ as follows: 
\begin {eqnarray}
j=\gamma{2\pi T\over{\cal I}^2({\cal E})}{\rm sh}{\pi {\cal E}\over T}, 
\label{b} 
\end {eqnarray} 
where one introduces the integral
\begin {eqnarray}
{\cal I}({\cal E})=\int\limits^{x_e+2\pi}_{x_e}
\exp\left\{{F(x)\over T}\right\}{\rm d}x,\quad 
\sin x_e\equiv {{\cal E}\over\omega_0^2}, 
\label{c} 
\end {eqnarray} 
$T$ is the temperature. 
In ergodic systems, the diffusion coefficient determined by equality 
\begin {eqnarray}
\langle(x-jt)^2\rangle\equiv Dt
\label{d} 
\end {eqnarray} 
takes the form
\begin {eqnarray}
D\equiv T{\partial j\over \partial {\cal E}}=
\gamma{2\pi^2 T\over{\cal I}^2({\cal E})}{\rm ch}{\pi {\cal E}\over T}.
\label{e} 
\end {eqnarray} 
However, non-ergodicity effects arrive at  much more complicated form of
the diffusion coefficient \cite{6} 
\begin {eqnarray}
D=j\left[\pi{\rm cth}{\pi {\cal E}\over T}+
{1\over 2}{\rm arcsin}{{\cal E}\over\omega_0^2}+
{{\cal I}{\cal E}\over 2(\omega_0^4-{\cal E}^2)}\right], 
\label{f} 
\end {eqnarray} 
where ${\cal I}$ is characteristic magnitude of the integral (\ref{c}). 
Dependencies $j({\cal E})$, $D({\cal E})$ related 
to Eqs.(\ref{b}), (\ref{f}) are depicted 
in Fig.~8. 

\section{Nonlinear resonance conditions}

With accounting dissipation effects shifting the bare frequency 
$\omega_0$ according to Eq.(\ref{22a}), the resonance condition reads
\begin{equation}
m_n\varpi=\Omega,\quad  
\varpi\equiv\sqrt{[\omega({\cal H})]^2-2^{-2}}, \quad n=1, 2,\dots,
\label{47}
\end{equation}
where $\Omega$ is the frequency of external signal, 
$\omega({\cal H})$ is the energy-dependent proper frequency 
determined by Eqs.(\ref{36}), (\ref{35}), 
$m_n$ is a resonance multiplicity being 
a rational number related to the natural one $n$. 
Formally, this condition 
means that we have to consider the phase portrait plane,  
which is revolved with angle velocity $m_n\varpi$. 
Then, the external addition of the Hamiltonian (\ref{17a})
is written in the form 
\begin{eqnarray}
{\cal H}_1=-Af_{00}\cos\vartheta,\quad 
\vartheta\equiv\left(\Omega-m_n\varpi\right)t,
\label{48}
\end{eqnarray}
where the internal power $f_{00}$ corresponds 
to resonance conditions (\ref{47}). 
Respectively, the zero term of the Hamiltonian (\ref{17a}) 
can be expanded near the resonant action $I_0$ as follows:
\begin{eqnarray}
{\cal H}_0\approx{\cal H}_{00} + \omega_{00}~\delta I + 
{\Delta_0\over 4}(\delta I)^2,\quad\delta I\equiv I-I_0.
\label{49}
\end{eqnarray}
Here resonant action $I_0$, as well as other resonant values 
\begin{eqnarray}
{\cal H}_{00}\equiv{\cal H}_0(I=I_0),\
\omega_{00}\equiv\left.{\frac{{\rm d}{\cal H}_0}{{\rm d}I}}\right|_{I=I_0},\
\Delta_0\equiv 2\left.{\frac{{\rm d}\omega}{{\rm d}I}}\right|_{I=I_0}=
2{\omega_{00}\over I_0}\alpha_0,\ \alpha_0\equiv\alpha(I=I_0) 
\label{50}
\end{eqnarray}
are given by resonance condition (\ref{47}). It is worthwhile to note 
that curvature $\Delta/2$ is determined by nonlinearity parameter 
(\ref{44}) taken at resonance condition $I=I_0$ (see Figs.5).
The energy-dependence of the curvature is characterized by the 
following asymptotics:
\begin{eqnarray}
\Delta\approx\pi^2\left\{\matrix{
(2\pi)^{-2}\left(1-{1\over 4}{{\cal H}\over{\cal H}_c}\right)
&\quad {\rm at} \quad{\cal H}\ll{\cal H}_c,\cr
\left(1-{{\cal H}\over{\cal H}_c}\right)^{-1}
\left(\ln{16\over 1-{\cal H}/{\cal H}_c}\right)^{-3}
\left[1-{1-{\cal H}/{\cal H}_c\over 4}
\ln{16\over 1-{\cal H}/{\cal H}_c}\right]\
&{\rm at} \ {\cal H}_c-{\cal H}\ll{\cal H}_c,\cr
\left({{\cal H}\over{\cal H}_c}-1\right)^{-1}
\left(\ln{16\over {\cal H}/{\cal H}_c-1}\right)^{-3}
\left[1+{{\cal H}/{\cal H}_c-1\over 4}
\ln{16\over{\cal H}/{\cal H}_c-1}\right]\
&{\rm at} \ {\cal H}-{\cal H}_c\ll{\cal H}_c,\cr
{1\over 2\pi^2}\left(1-{{\cal H}_c\over{\cal H}}\right)
&\quad {\rm at} \quad{\cal H}\gg{\cal H}_c.\cr}\right.
\label{50a}
\end{eqnarray}

As a result, Hamilton equations (\ref{20}) arrive at the equation 
type of Eq.(\ref{28}) for nonliniar pendulum:  
\begin{equation}
\ddot \vartheta + \omega_m^2\sin\vartheta = 0.  
\label{51}
\end{equation}
Here, instead of the proper frequency $\omega_0$,  
value $\omega_m$ stands for modulation frequency being determined as follows:
\begin{equation}
\omega_m=\left({A\Delta f_{00}\over 2}\right)^{1/2}.
\label{52}
\end{equation}
Respectively, maximum value of the resonant energy variation 
and corresponding magnitude for the action 
\begin{equation}
\delta{\cal H}_m=Af_{00},\quad
\delta I_m=\left({4Af_{00}\over\Delta}\right)^{1/2}
\label{53}
\end{equation}
are determined by Eqs.(\ref{48}), (\ref{49}) 
to characterize a resonance window.

Thus, we obtain the following picture of nonlinear resonance. 
At given magnitudes of the frequency $\Omega$ of external signal, 
the condition (\ref{47}) fixes the system energy ${\cal H}$ as follows: 
\begin{equation}
{\pi\over\sqrt{\omega_0^{-2}+(2\Omega/m_n\omega_0)^2}}=K({\kappa})
\left\{\matrix{1&\qquad {\rm at}\qquad \kappa\leq 1,\cr
\kappa^{-1}&\qquad {\rm at}\qquad \kappa\geq 1.\cr}\right.
\label{54}
\end{equation}
According to Fig.~9, the corresponding dependence ${\cal H}_{00}(\Omega)$ 
has two branches, the lower of which relates to finite motion, 
the upper -- to infinite one. 
Energy ${\cal H}_{00}$ related to the former falls down monotonously 
from the upper magnitude fixed by condition 
$K({\kappa})=\pi\omega_0$ to zero 
within interval $0<\Omega<\Omega_{m}$, where maximum frequency is  
\begin{equation}
\Omega_{m}=m_n\omega_0\sqrt{1-(2\omega_0)^{-2}}.
\label{55}
\end{equation}
Respectively, energy of the infinite motion arises monotonously 
from minimal magnitude fixed at $\Omega=0$ by condition 
$\kappa^{-1}K(\kappa^{-1})=\pi\omega_0$ 
to ${\cal H}_{00}\to\infty$ at $\Omega\to\infty$.
Using abbreviated notation
\begin{equation}
\tilde\Omega^2\equiv 1+(2/m_n)^2\Omega^2,
\label{55a}
\end{equation}
one obtains the following asymptotics: 
\begin{eqnarray}
{\cal H}_{00}\approx{\cal H}_c\left\{\matrix{
1 - 16\exp\left[-2\pi(\omega_0/\tilde\Omega)\right]
&\quad{\rm at}\quad \Omega\ll\Omega_m,\cr
4\left[1-(2\omega_0)^{-2}\right]{\Omega_m-\Omega\over\Omega_m}
&\quad{\rm at}\quad 0<\Omega_m-\Omega\ll\Omega_m\cr}\right.
\label{56}
\end{eqnarray}
for finite motion and
\begin{eqnarray}
{\cal H}_{00}\approx{\cal H}_c\left\{\matrix{
1 + 16\exp\left[-2\pi(\omega_0/\tilde\Omega)\right]
&\quad{\rm at}\quad \Omega\ll\Omega_m,\cr
\left({\tilde\Omega\over 2\omega_0}\right)^{2}
&\quad{\rm at}\quad\Omega\gg\Omega_m\cr}\right.
\label{56a}
\end{eqnarray}
for infinite one. 
Frequency-dependencies of the resonant magnitudes 
$I_0(\Omega)$, $\omega_{00}(\Omega)$, $\alpha(\Omega)$, 
$\Delta_0(\Omega)$ of the action, the proper frequency, 
the nonlinearity parameter and the double curvature of curve ${\cal H}_0(I)$ 
are depicted in Figs. 10, 11.  
In the case of the finite motion, these dependencies are characterized 
by the following asymptotics:
\begin{eqnarray}
I_0\approx8\omega_0\left\{\matrix{\pi^{-1}
\left\{1-8\pi(\omega_0/\tilde\Omega)
\exp\left[-2\pi(\omega_0/\tilde\Omega)\right]\right\}
&\quad{\rm at}\quad \Omega\ll\Omega_m,\cr
\left[1-(2\omega_0)^{-2}\right]{\Omega_m-\Omega\over\Omega_m}
&\quad{\rm at}\quad 0<\Omega_m-\Omega\ll\Omega_m;\cr}\right.
\label{57}
\end{eqnarray}
\begin{eqnarray}
\omega_{00}\approx\omega_0\left\{\matrix{
{\tilde\Omega\over 2\omega_0}
&\quad{\rm at}\quad \Omega\ll\Omega_m,\cr
1-\left[1-(2\omega_0)^{-2}\right]{\Omega_m-\Omega\over\Omega_m}
&\quad{\rm at}\quad 0<\Omega_m-\Omega\ll\Omega_m;\cr}\right.
\label{58}
\end{eqnarray}
\begin{eqnarray}
\Delta_{0}\approx{1\over 4}\left\{\matrix{
\left({\tilde\Omega\over 2\omega_0}\right)^2
\left\{ { \tilde\Omega\over 8\pi\omega_0}
\exp\left[2\pi(\omega_0/\tilde\Omega)\right]-1\right\}
&\quad{\rm at}\quad \Omega\ll\Omega_m,\cr
1-\left[1-(2\omega_0)^{-2}\right]
{\Omega_m-\Omega\over\Omega_m}
&\quad{\rm at}\quad 0<\Omega_m-\Omega\ll\Omega_m.\cr}\right.
\label{59}
\end{eqnarray}
Respectively, for the infinite motion one has:
\begin{eqnarray}
I_0\approx2\omega_0\left\{\matrix{{4\over\pi}
\left\{1+8\pi(\omega_0/\tilde\Omega)
\exp\left[-2\pi(\omega_0/\tilde\Omega)\right]\right\}
&\quad{\rm at}\quad \Omega\ll\Omega_m,\cr
\tilde\Omega/\omega_0
&\quad{\rm at}\quad\Omega\gg\Omega_m;\cr}\right.
\label{57a}
\end{eqnarray}
\begin{eqnarray}
\omega_{00}\approx\tilde\Omega/2
&\quad{\rm at}\quad \Omega\ll\Omega_m
 \quad{\rm and}\quad\Omega\gg\Omega_m;
\label{58a}
\end{eqnarray}
\begin{eqnarray}
\Delta_{0}\approx{1\over 2}\left\{\matrix{
{1\over 2}
\left({\tilde\Omega\over 2\omega_0}\right)^2
\left\{ { \tilde\Omega\over 8\pi\omega_0}
\exp\left[2\pi(\omega_0/\tilde\Omega)\right]+1\right\}
&\quad{\rm at}\quad \Omega\ll\Omega_m,\cr
1-(2\omega_0/\tilde\Omega)^2
&\quad{\rm at}\quad\Omega\gg\Omega_m.\cr}\right.
\label{59a}
\end{eqnarray}
At $\omega_0\gg 1/2$, when $\Omega_m\approx m_n\omega_0$,
effective frequency $\tilde\Omega$ may be replaced by the short-cut term 
$(2/m_n)\Omega\gg 1$.

If the oscillations were linear in character, 
the resonance might realize within window located between energies 
${\cal H}_{00}-\delta{\cal H}_m$ and ${\cal H}_{00}+\delta{\cal H}_m$. 
In this window, the oscillations would have single frequency $\varpi$ 
determined by Eq.(\ref{47}) and would be modulated 
with frequency $\omega_m$ given by Eq.(\ref{52}).
However, nonlinearity effects described in Subsection 3.2 
arrive at expanding the harmonics number to magnitude 
$N_0=\omega_0/\omega_{00}>1$ and narrowing the energy window to width  
${\cal H}_m\equiv 2\omega_m^2$. So, nonlinear resonance is realized 
for a share of nonlinear oscillations determined by the ratio 
\begin{equation}
{{\cal H}_m\over\delta{\cal H}_m}=\Delta_0=2{\omega_{00}\over I_0}\alpha_0, 
\label{60}
\end{equation}
where Eqs.(\ref{50}), (\ref{52}), (\ref{53}) are taken into account. 
The double curvature $\Delta(\Omega)$ 
decreases monotonously with the external frequency growth within the interval 
$0<\Omega<p\Omega_m$. This means that preference of the nonlinear resonance 
decreases with this frequency growth. 

\section{Stochastic resonance conditions}

The stochastic resonance is known to be observed at condition (\ref{47}), 
where $m_n=(2n)^{-1}$, $n=1, 2,\dots$. 
This condition means that during a period $T=2\pi/\Omega$ 
of the external oscillation the stochastic system has a time to overcome 
an energy barrier $\Delta F$ even times $2n$ \cite{5}.
The proper frequency of stochastic resonance $\omega\equiv 2\pi r_K$ 
is reduced to the Kramers' rate $r_K$ given by Eq.(\ref{A2}). 
This case differs from the above considered case of nonlinear pendulum 
by only inserting temperature $T$ instead of the system energy ${\cal H}$. 
In the case of double well potential considered in Subsection 3.3,  
the energy barrier $\Delta F=(\omega_0/2)^2$ is related to the temperature 
by characteristic parameter $\kappa^2\equiv 4T/\omega_0^2$.
Then, the frequency of the barier overcoming is defined by equation 
\begin{equation}
\omega=\omega_{K}\exp(-\kappa^{-2}),\quad
\kappa^2\equiv 4T/\omega_0^2
\label{62}
\end{equation}
instead of the corresponding equation (\ref{35}) for nonlinear pendulum. 
Taking into consideration dissipation effects, we obtaine the following 
condition of the stochastic resonance:   
\begin{equation}
T_0={\omega_0^2\over 2}
\left[\ln{(2\omega_{K})^2\over 1+(4n)^2\Omega^2}\right]^{-1}.
\label{63}
\end{equation}
As is shown in Fig.~12, the resonant temperature $T_0(\Omega)$ arises 
monotonously, taking indefinite values at characteristic frequency
\begin{equation}
\Omega_m={\omega_{K}\over 2n}\sqrt{1-(2\omega_{K})^{-2}}.
\label{64}
\end{equation}
Basing on the dependence $T_0(\Omega)$ replacing above used relation 
${\cal H}(\Omega)$, we are in position to consider the stochastic resonance conditions 
in analogy with nonlinear pendulum.  

Here, instead of Eqs.(\ref{32}), (\ref{34}), (\ref{35}), 
the equalities (\ref{32b}), (\ref{34c}), (\ref{35c}) 
determine the resonant magnitudes $I_0$, $\omega_{00}$, $\alpha$, 
$\Delta$ of the action,  
the proper frequency, the nonlinearity parameter and the double 
curvature of the resonant dependence ${\cal H}_{00}(I)$.  
Corresponding dependencies on the temperature are depicted in Figs. 13, 14 
to show indefinite increase of the proper frequency and the action  
with tending to the characteristic magnitude (\ref{64}). 
Related values of the nonlinearity parameter and the curvature decrease thereby.

\section{Conclusion}

The observed picture of nonlinear oscillations shows that with energy increase
the single harmonic oscillation transforms to a harmonics superposition, 
whose number increases monotonously to infinity, whereas frequency 
and amplitude decrease to zero.
In other words, with tending to a threshold energy, 
transition of the harmonic oscillations into a set of solitons is observed. 
We have shown that such behaviour, being typical for both the nonlinear pendulum 
and the double well potential, is characterized by nonlinearity parameter (\ref{44}). 

Our choice of the action $S$ and angle $\vartheta$ as principle variables 
is caused by that Hamiltonian of free nonlinear oscillations 
does not depend on the angle $\vartheta$.
In accordance with Eq.(\ref{51}), switching on external harmonic signal 
arrives at a modulation of nonlinear oscillations with  
characteristic frequency (\ref{52}) and nonlinearity parameter (\ref{50}). 
It is appeared that nonlinearity effect narrows the resonance window 
to width fixed by the ratio (\ref{60}) that is reduced 
to the curvature $\Delta$ of the dependence ${\cal H}(I)$. 
According to the dependence $\Delta_0(\Omega)$ depicted in Fig. 14, this window is
shrunken with the external frequency growth. 
As a result, preference of the stochastic resonance 
decreases with growth of the frequency $\Omega$ of external signal. 

\section*{Acknowledgments}\label{sec:level1}

I am grateful to Anton Tiptyuk for the picturing figures.

\newpage

\begin{center}
{\bf Figure captions}
\end{center}

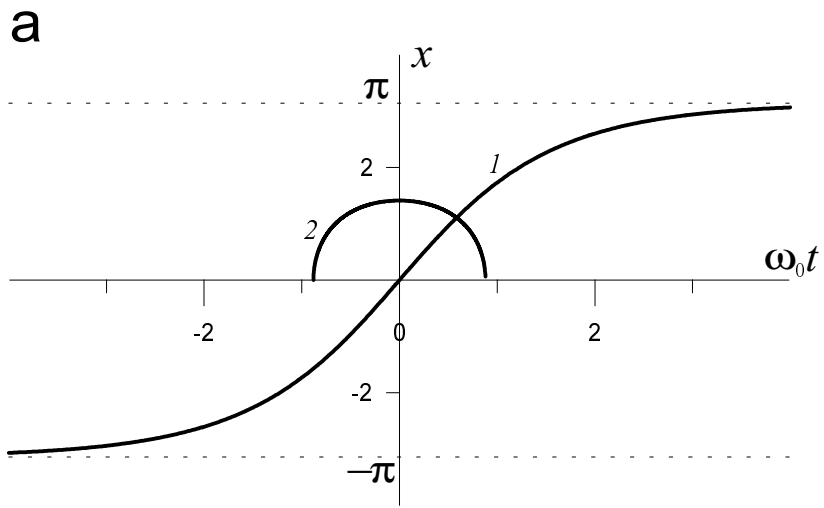
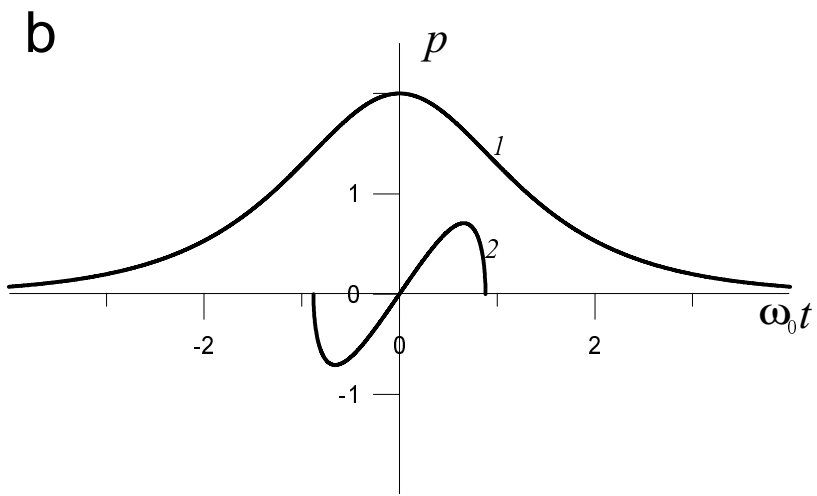
Fig.~2. (a) Motion laws for nonlinear pendulum (curve 1) and double well potential (curve 2); 
(b) Corresponding time dependencies of the conjugate momentum.

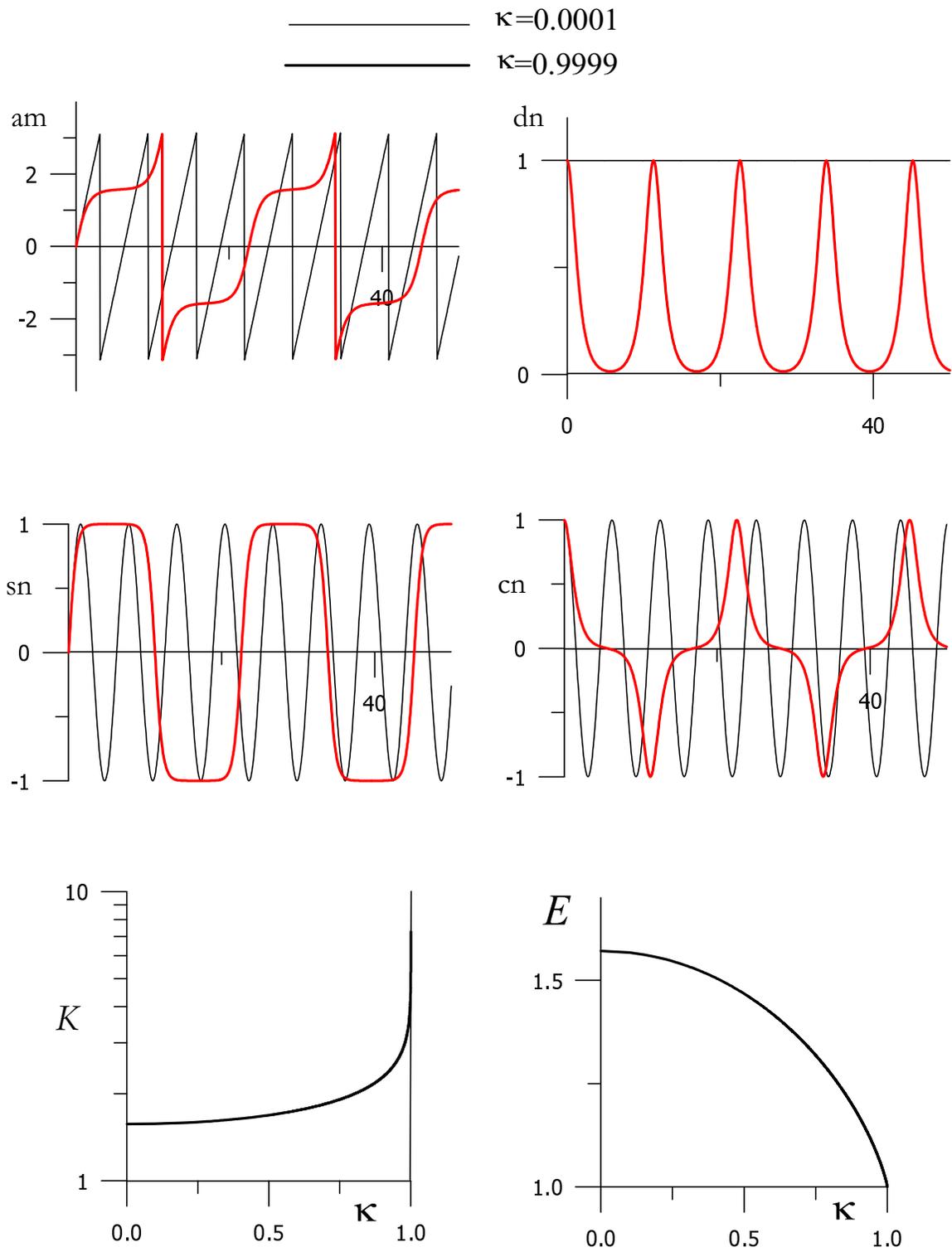
Fig.~3. Form of Jacobian elliptic functions and integrals. 

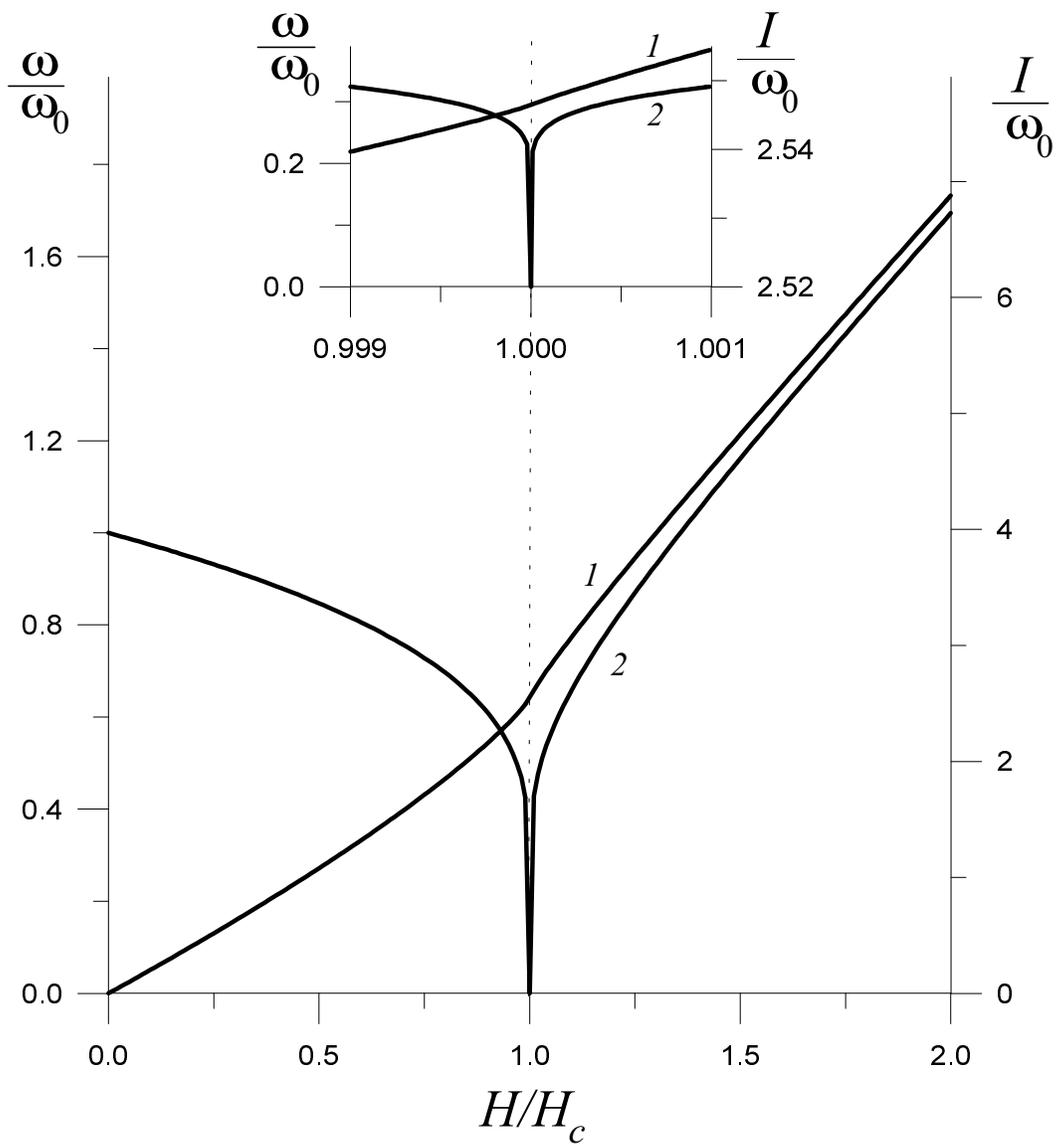
Fig.~4. Energy dependencies of the action and the proper frequency for nonlinear pendulum. 

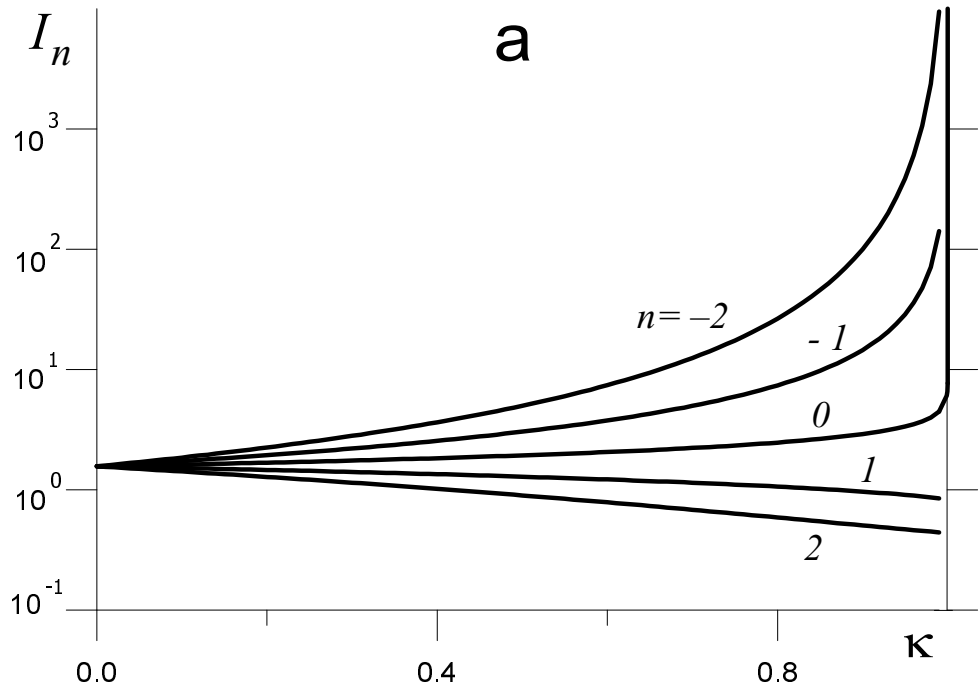
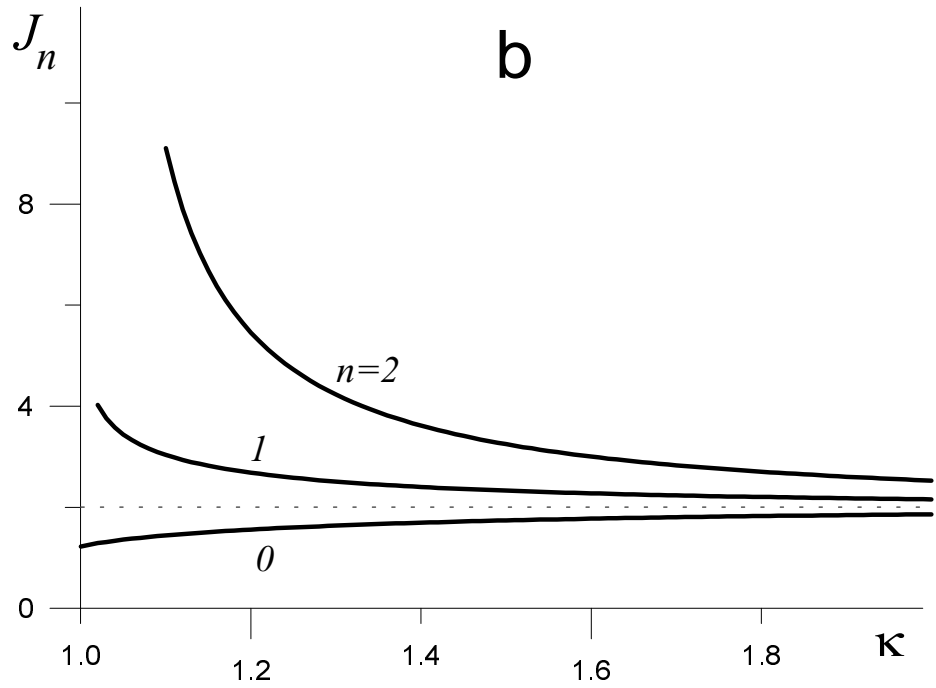
Fig.~6. Form of the integrals (\ref{33a}). 

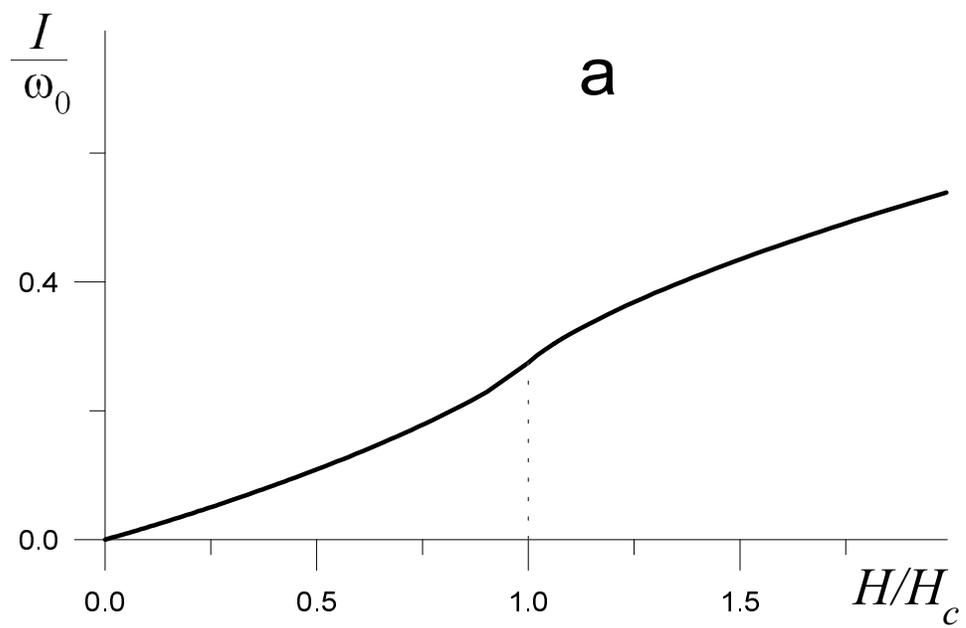
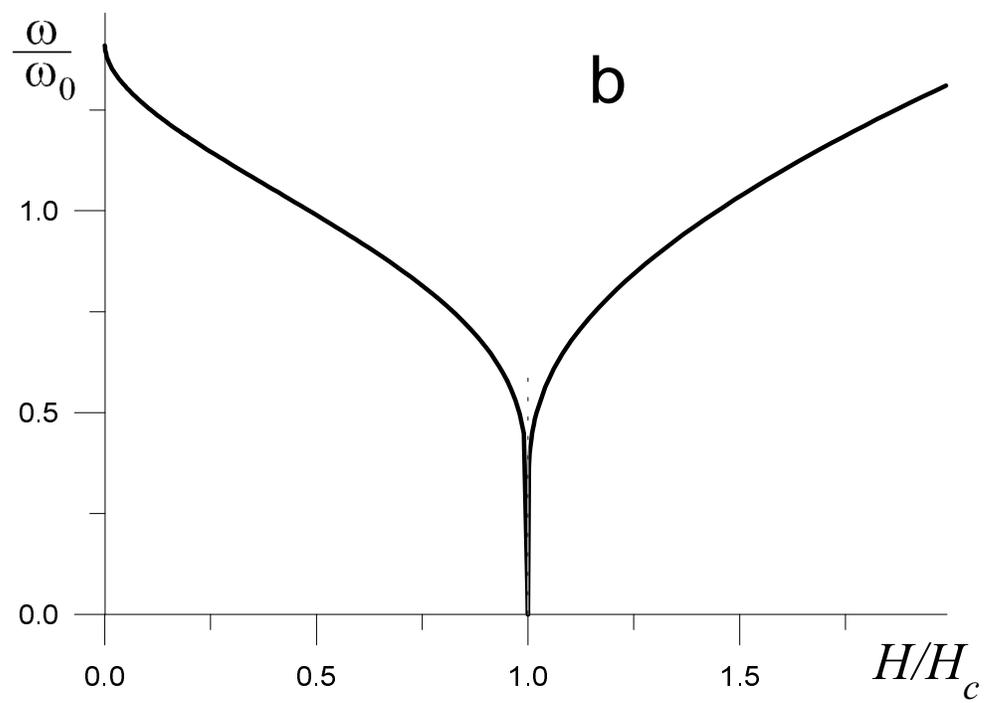
Fig.~7. Energy dependencies of the action (a) and the proper frequency (b) for double well potential. 

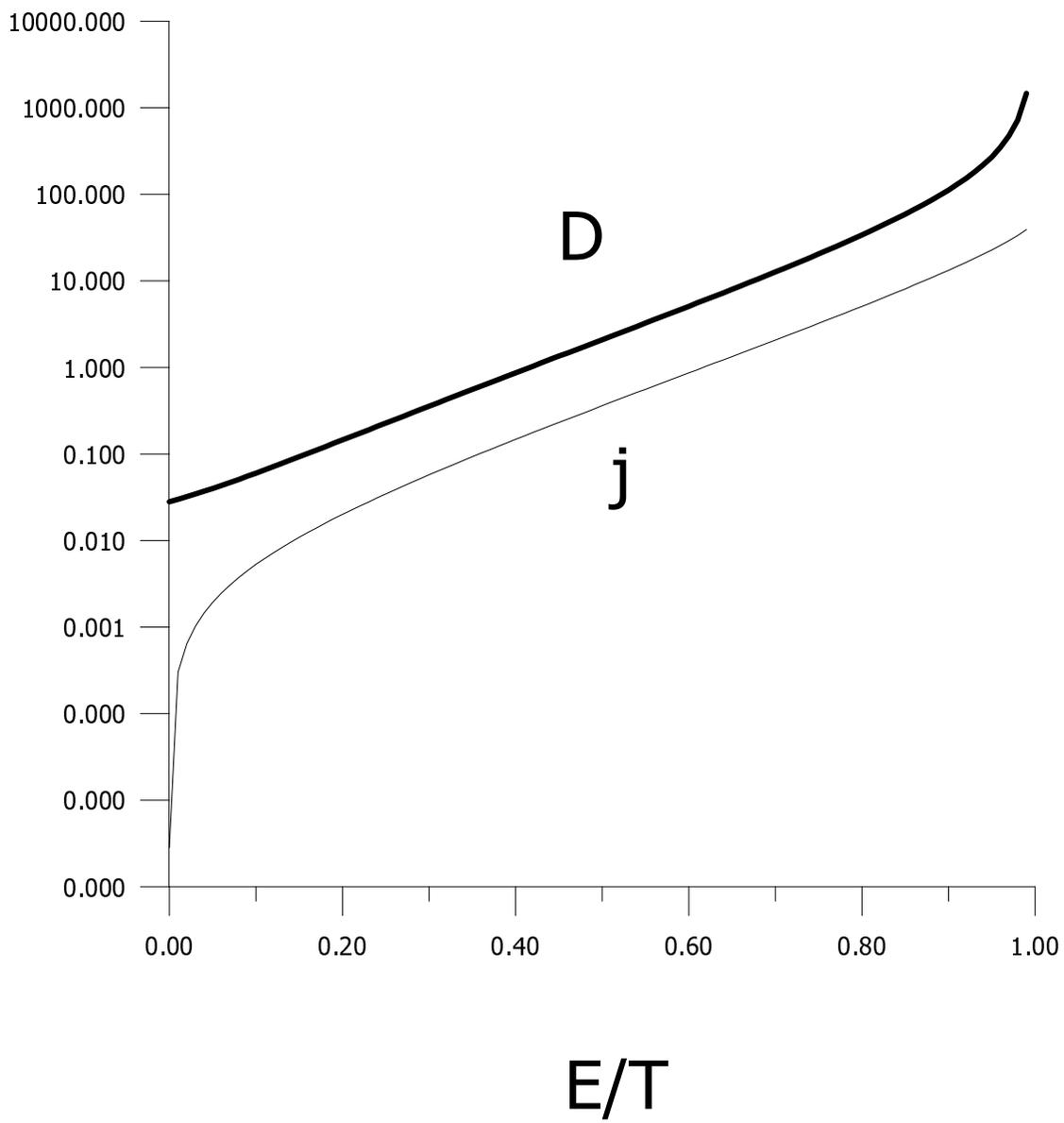
Fig.~8. Dependence of the flux $j$ and the diffusion coefficient $D$ on the force $E$ 
of external field for nonlinear pendulum. 

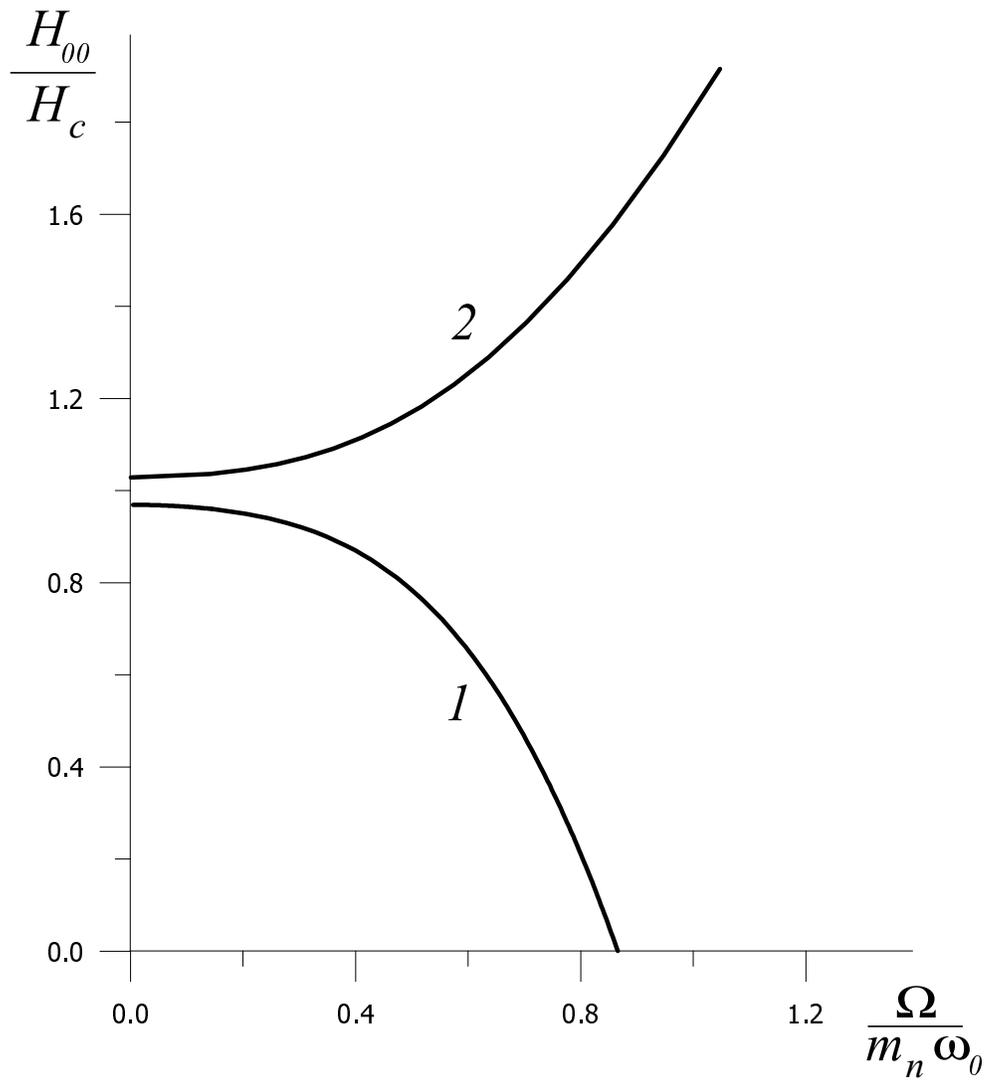
Fig.~9. Dependence of the nonlinear pendulum energy on the external signal frequency: 
curve 1 relates to finite motion, curve 2 -- to infinite one. 

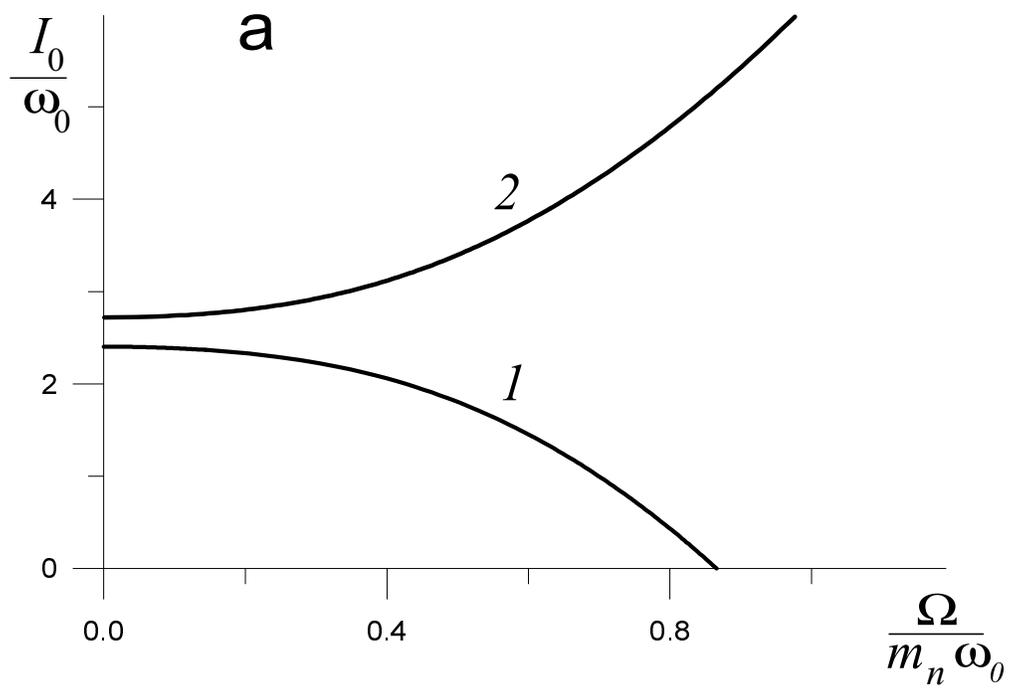
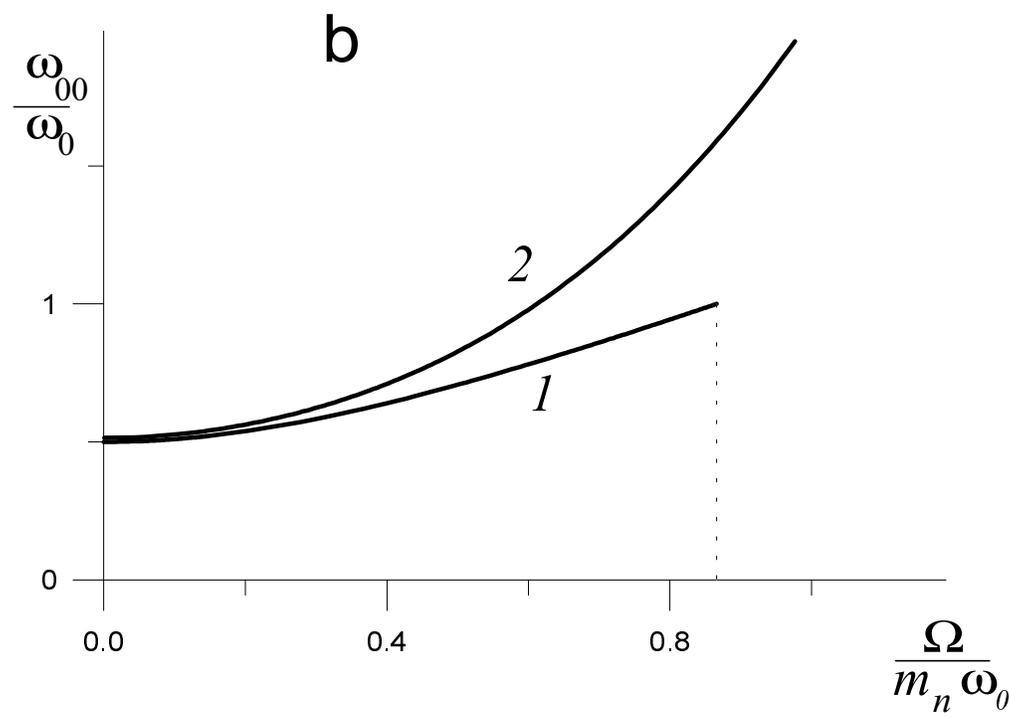
Fig.~10. Frequency dependencies of the action (a) and the proper frequency (b) for nonlinear pendulum: 
curve 1 relates to finite motion, curve 2 -- to infinite one. 

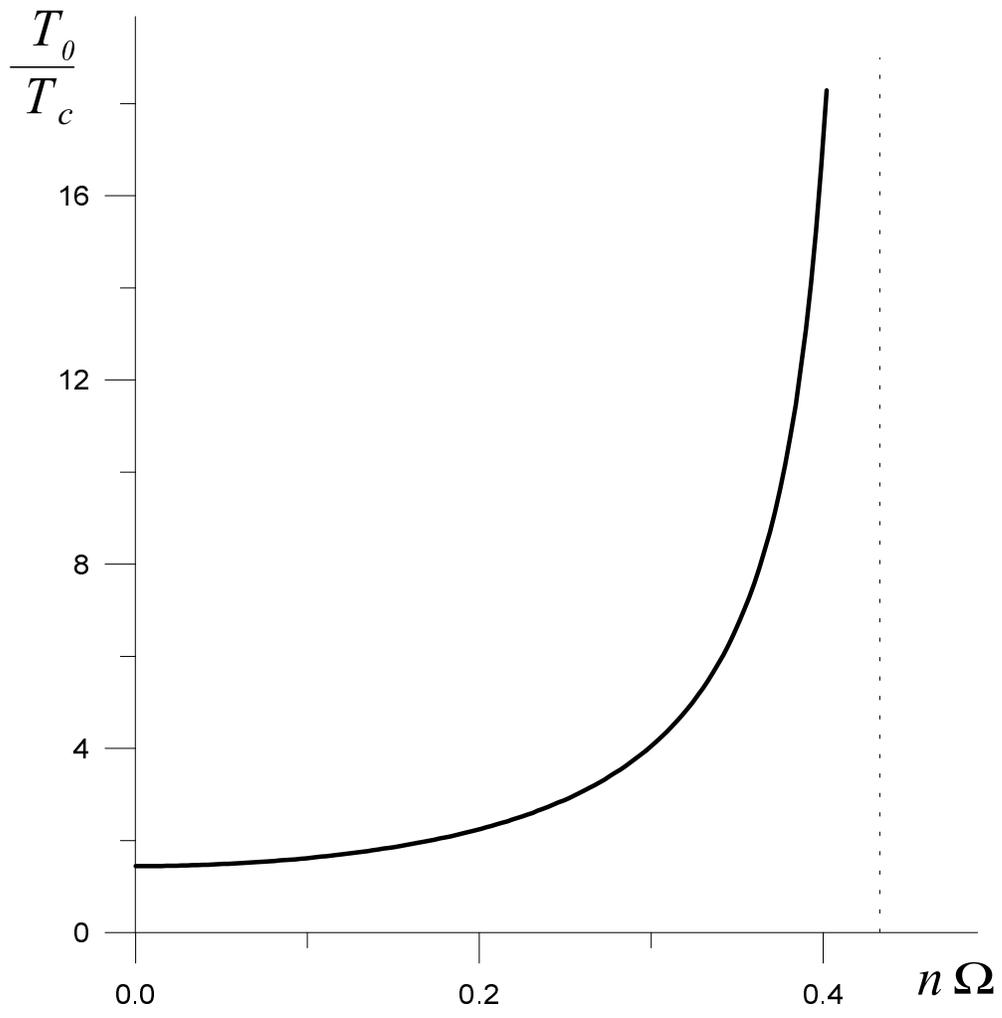
Fig.~12. Dependence of the stochastic resonance temperature on the external signal frequency. 

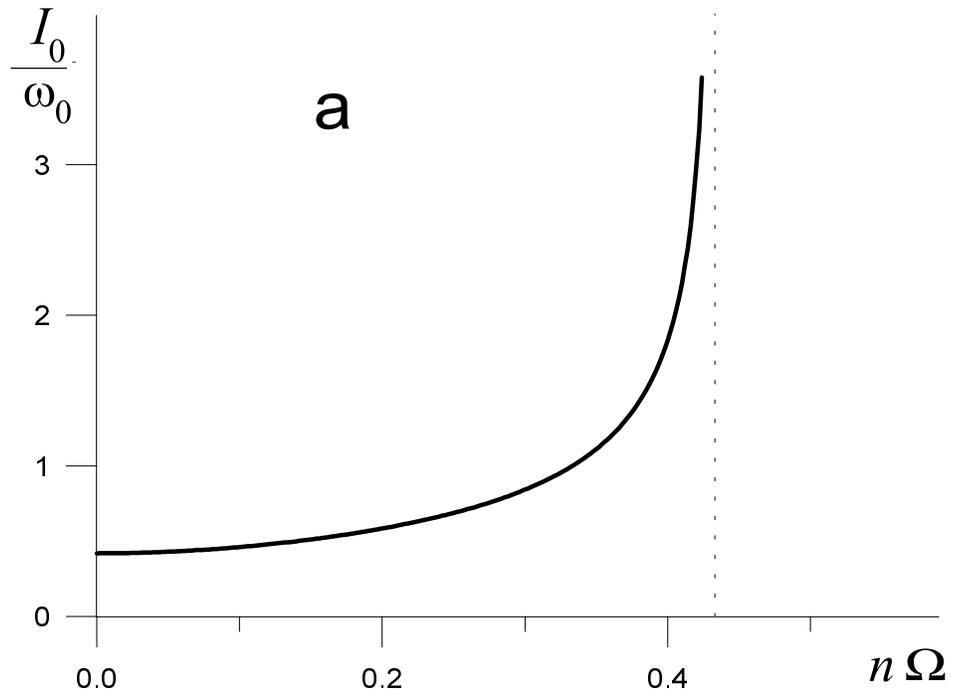
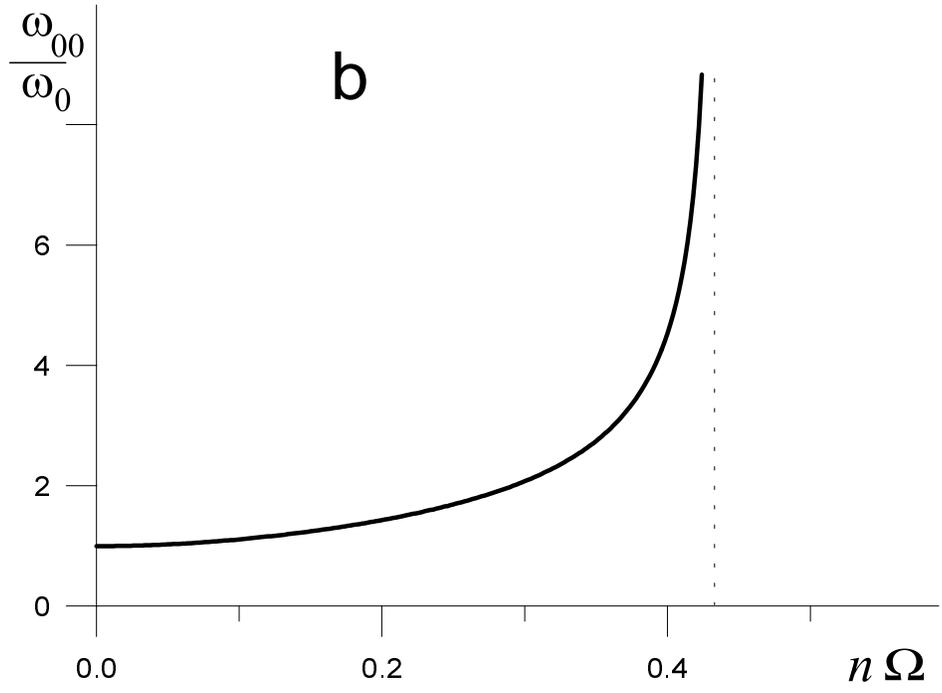
Fig.~13. Frequency dependencies of the action (a) and the proper frequency (b) of the stochastic resonance.

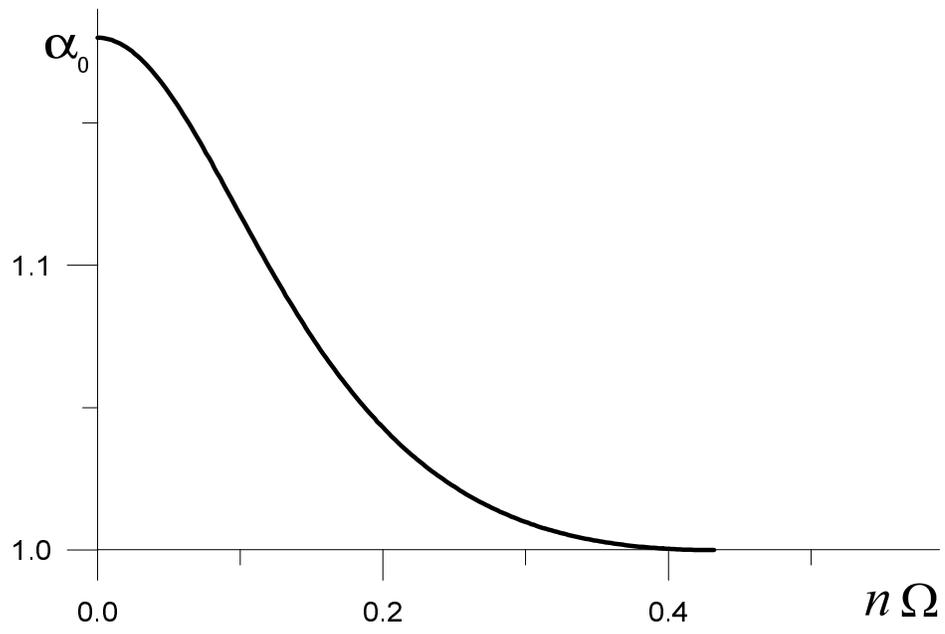
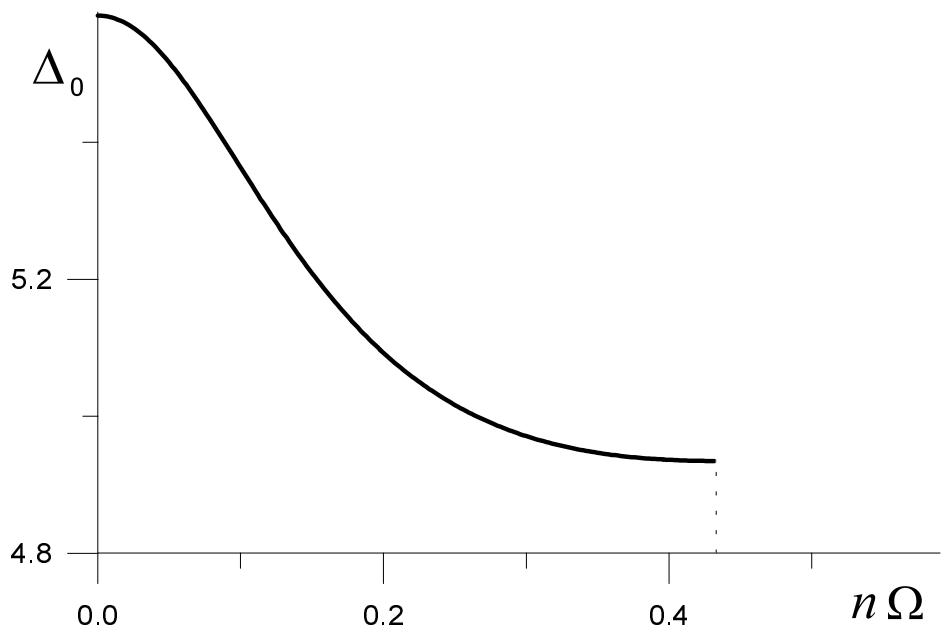
Fig.~14. Frequency dependencies of the nonlinearity parameter $\alpha_0$ and curvature $\Delta_0$ 
of the dependence ${\cal H}(I)$ at stochastic resonance. 

\end{document}